\documentstyle[12pt,epsfig]{article}
\begin{document}
\newcommand {\be}{\begin{equation}}
\newcommand {\ee}{\end{equation}}
\newcommand {\ba}{\begin{eqnarray}}
\newcommand {\ea}{\end{eqnarray}}
\newcommand {\bea}{\begin{array}}
\newcommand {\cl}{\centerline}
\renewcommand {\thefootnote}{\fnsymbol{footnote}}
\newcommand {\eea}{\end{array}}
\vskip .5cm

\baselineskip 0.65 cm
\begin{flushright}
$\hspace {2cm}$\\
hep-ph/0105105 \\
\today
\end{flushright}
\begin{center}
{\Large{\bf Neutrino Mass Spectrum and  Future Beta Decay
Experiments}} \\

\vskip 0.5cm

{Y.\, Farzan~$^a$  \footnote{ E-mail:farzan@sissa.it} , O.\,L.\,G.
\, Peres~$^{b,c}$
\footnote{E-mail:orlando@ifi.unicamp.br} and
A.\, Yu.\,
Smirnov~$^{c,d}$  \footnote{E-mail:smirnov@ictp.trieste.it} \\}

\vskip 0.3cm

$^a$ {\it Scuola Internazionale superiore di Studi Avanzati \\ via Beirut 4,
I,34014 Trieste, Italy \\}
$^b${\it Instituto de F\' {\i}sica Gleb Wataghin,
    Universidade Estadual de Campinas, UNICAMP\\
    13083-970 Campinas SP, Brazil\\}
$^c$ {\it The Abdus Salam International Centre for Theoretical Physics,
I-34100 Trieste,Italy}\\
$^d$ {\it Fermi National Accelerator Laboratory,
P. O. Box 500, Batavia, IL 60510}
\end{center}
 \renewcommand{\baselinestretch}{1.0}

\begin{abstract}
We study  the discovery potential of future
beta decay experiments
on searches for the neutrino mass in the sub-eV
range, and,  in particular, KATRIN experiment
with sensitivity $m > 0.3$ eV.
Effects of neutrino mass and mixing on the beta decay
spectrum in the neutrino schemes which
explain the solar and atmospheric neutrino data are discussed.
The schemes which lead to  observable effects contain
one or two  sets of  quasi-degenerate states.
Future beta decay measurements will allow to check the three neutrino
scheme with mass degeneracy, moreover, the possibility appears to measure
the CP-violating Majorana phase.
Effects in the four neutrino schemes which can also
explain the LSND  data  are strongly restricted  by the results of Bugey
and CHOOZ oscillation experiments: Apart from bending of the
spectrum and the shift of the end point
one expects   appearance of small
kink of ($ < 2\% $) size or suppressed  tail after bending of the
spectrum with rate below  2 \% of the expected rate for zero neutrino
mass. We consider  possible implications of future beta decay
experiments for the neutrino mass spectrum, the  determination of the
absolute scale of neutrino mass and for  establishing
the nature of neutrinos.
We show that beta decay measurements in combination with
data from the oscillation and double beta decay experiments
will allow to establish the structure of the
scheme (hierarchical or non-hierarchical), the type of the hierarchy
or ordering of states (normal or inverted) and to
measure the relative CP-violating phase in the solar pair of states.
\end{abstract}
\renewcommand{\baselinestretch}{2.0}
\vskip 0.1cm

\noindent
{\it PACS:} 14.60.Pq; 14.60.Lm\\
{\it Keywords:} Neutrino masses and mixing; Beta decay


\newpage
\section{Introduction}

The reconstruction of the neutrino mass spectrum is one of the
fundamental problems of particle physics.
The program includes the determination of the number of mass eigenstates,
and of the values of masses, mixing parameters and CP-violating
phases.

At present,  the evidence for  non-zero neutrino
mass follows from oscillation experiments which  allow
to measure  the mixing parameters
$|U_{\alpha j}|$, the mass squared differences  and, in principle, the so
called Dirac
CP-violating phases. However, the absolute values of  the neutrino masses
cannot be determined.  From the  oscillation
experiments  one  can only extract  a  {\it  lower} bound
on the absolute value of neutrino mass. Obviously,
for a given   $\Delta m^2$, at least  one of the mass  eigenvalues should
satisfy inequality:
$$
m_i \geq \sqrt{|\Delta m^2|}.
$$
Thus, the oscillation interpretation of the atmospheric neutrino
data~
\cite{atm}  gives the bound:
$$
m_3 \geq \sqrt{\Delta m^2_{atm}} \sim (0.04 -  0.07) ~ {\rm eV}.
$$

Clearly, without knowledge of the absolute values of neutrino masses
our picture of Nature at quark-lepton level will be incomplete.
The knowledge of absolute values of
neutrino masses is crucial for  understanding the  origin of the
fermion masses in general, the quark-lepton
symmetry and  unification.
The determination of the absolute mass scale of neutrinos
is at least as important as
the determination of  other fundamental parameters such as
the CP-violating phases and the mixing angles. Actually, it may have
even more
significant and straightforward implications for the fundamental theory.
It is the absolute mass which determines the scale of new
physics.

The absolute values of masses
have crucial implications for
astrophysics and cosmology, in particular, for
structure formation in  the Universe.
In fact, the recent analysis of the latest CMB data
(including BOOMERanG, DASI, Maxima
and CBI), both alone and jointly with other cosmological data
({\it e.g.},  galaxy clustering and the Lyman Alpha Forest) shows that
\cite{recent}
\be
m_\nu <2.2 \ \ {\rm eV}, \label{recent}
\ee
for a single neutrino in eV range.
Future observations can improve this bound.
The Planck experiment will be sensitive to  neutrino masses down to
 $m_{\nu} \sim 1$ eV \cite{planck}.
However, the cosmological data may not be conclusive.  Even if
some effects are found, it will be difficult  to  identify their
origin. Modification of the original spectrum of the density
fluctuations can mimic to some extent the neutrino mass effect.
If no distortion is
observed in the spectrum,  one can put an  upper bound on the neutrino
mass  assuming, however, that there is
no conspiracy which leads to cancellation of different effects
\cite{conspiracy}.
Therefore independent measurements  of the neutrino mass are needed
and their results will be used in the
analysis of the cosmological data as an input
deduced from particle physics.

Several methods have been  proposed to determine neutrino masses by using
the
supernova neutrino data. One method is based on searches for the energy
ordering of
events which has, however, rather low sensitivity \cite{sato}.
The limits on the mass can be also obtained from observations of
sharp time structures in the signals.
It was suggested to study the time
distribution of detected  neutrino events emitted from supernova which
entails to black
hole formation \cite{Beacom}. By this
method
Super-Kamiokande can measure values of  the  $\nu_e$ mass down to 1.8
eV and SNO can put  an  upper  bound 20 eV
on the $\nu_\mu$ and $\nu_\tau$ masses \cite{Beacom} .
(Clearly this bound on the $\nu_\mu$ and $\nu_\tau$ masses is much weaker
than bounds implied  by
combined analysis of the solar and  atmospheric neutrino data and direct
measurements of the $\nu_e$ mass.)
In this case one   can check the still non-excluded possibility
in which  the solar neutrino problem is solved by the oscillations to
sterile
neutrino
and the masses of $\nu_\mu$ and $\nu_\tau$ are in 20 eV range.
(Such neutrinos should be unstable in cosmological time.)
The absolute values of the neutrino masses can be determined in
the assumption
that the cosmic rays with energies above the GZK cutoff are produced
in annihilation of the ultra-high energy neutrinos with the cosmological
relic neutrinos \cite{fargion, burst, Pas}. The analysis of the observed
energy
spectrum of cosmic rays above $10^{20}$ eV gives  the mass
$m_{\nu} = (1.5 - 3.6)$ eV, if the power-like part of the ultra-high
energy
cosmic rays spectrum is produced in
Galactic halo, and $m_{\nu} = (0.12  - 0.46)$ eV, if this part  has the
extragalactic origin \cite{fodor}.

Neutrinoless double beta decay ($2 \beta 0 \nu$) searches are sensitive
to the Majorana mass of the electron neutrino.
However,  in  the presence of mixing the situation can be  rather
complicated:
The
{\it effective} Majorana  mass of $\nu_e$ relevant for the
$ 2 \beta 0 \nu$-decay, $m_{ee}$, is a combination of mass
eigenvalues and mixing parameters given by
\be
m_{ee}=\left| \ \ \sum_i m_i U_{ei}^2 \ \ \right| \label{mee}.
\label{m_ee}
\ee
{}From this expression it is easy to find  that if the  $2\beta 0
\nu$-decay  is discovered with the  rate which corresponds to
$m_{ee}$, at least  one of the mass eigenvalues should satisfy
the inequality \cite{onethird}
\be
m_i \geq {m_{ee} \over n},
\ee
where $n$  is the number of neutrino mass eigenstates that mix in
the electron neutrino.
This bound is based on the assumption
that exchange of the light Majorana neutrinos is the only mechanism of
the $2\beta 0 \nu$-decay and all other  possible contributions are absent or
negligible.
Another uncertainty is related to $n$.  We know only the lower bound:
$n
\geq 3$.

The best present bound on the $2 \beta 0 \nu$-decay obtained by
Heidelberg-Moscow group gives \cite{newbound}
\be
\label{dbbound}
m_{ee} < 0.34 \ \  (0.26 )\  \ {\rm eV}, \ \ \
{\rm  \ \  \ \   90~\% ~~ (68 \%)  \ \ C.L.}.
\ee
This bound, however, does not include systematic errors related to nuclear
matrix elements~\footnote{In what follows we will use the bound
(\ref{dbbound}) in  our estimations for definiteness. At the same time,
one should keep in mind that due to uncertanties of nuclear matrix
element
the values of $m_{ee}$  up to $\sim 0.5$ eV can not be excluded.}.

A series of new experiments is  planned with increasing sensitivity to
$m_{ee}$: CUORICINO \cite{cino}, CUORE ($m_{ee} \sim 0.1$ eV) \cite{cuore},
MOON ($m_{ee} \sim 0.03$
eV)
\cite{moon} and GENIUS ($m_{ee} \sim 0.002$ eV) \cite{genius}.

Although the knowledge of $m_{ee}$   provides
 information on the mass
spectrum independent of $\Delta m^2$'s,
from $m_{ee}$ one cannot  infer
the absolute values of neutrino masses without additional assumptions.
Since in general the mixing elements  are complex there may be a
strong
cancellation in the sum (\ref{m_ee}). Moreover, to
induce the  $2 \beta 0 \nu$ decay, $\nu_e$ must
be a Majorana particle.

The information about the absolute values of the masses can be extracted
from
kinematic studies of reactions in which a neutrino or an anti-neutrino is
involved ({\it e.g.}, beta decays or lepton capture). The most sensitive
method
for
this purpose is  the study of the electron spectrum in the tritium decay:
\be
{\rm ^3H \rightarrow
^3He+e^-+\stackrel {-} \nu_e.}
\label{decay}
\ee
In absence of mixing,
the energy spectrum of $e^-$ in (\ref{decay}) is described  by
\begin{equation}
{dN \over dE} = R(E)[(E-E_0)^2-m_\nu^2]^{1 \over 2},
\label{mother}
\end{equation}
(see, {\it e.g.}, \cite{ketab}) where $E$ is the energy of electron, $E_0$
is the
total decay
energy and $R(E)$ is a $m_\nu$-independent function given by
\be
R(E)=G_F^2 {m_e^5  \over 2 \pi^3} \cos^2 \theta_C
|M|^2 F(Z,E)p E(E_0-E)~.
\label{re}
\ee
Here $G_F$ is the Fermi constant, $p$ is the momentum of the
electron, $\theta_C$ is the Cabibbo angle and $M$ is the nuclear matrix
element.
$F(Z,E)$ is a smooth  function of energy which describes
the  interaction of the produced electron in final state.
Both  $M$ and $F(Z,E)$ are independent of $m_\nu$, and the dependence of
the spectrum shape on $m_\nu$ follows  from the phase volume factor only.
The  bound on neutrino mass
imposed by the shape of the spectrum is independent of whether neutrino is
a Majorana or a Dirac particle.

The best present bound on the electron neutrino mass,
(obtained in  the assumption of no  mixing) is given by
Mainz tritium beta decay experiment \cite{mainz}:
\be
\label{mainz}
m_{\nu_e} \le 2.2 \ \ {\rm eV}  \ \ \ \ ({\rm 95 \%  \ \ C.L.}).
\ee
Analysis of the Troitsk results  leads to the ``conditional"
(after subtraction of the
excess of events near the end point)  bound
\cite{troitsk}
\be
m_{\nu_e} \le 2.5 \ \ {\rm eV}  \ \ \ \ ({\rm 95 \% \ \
C.L.}).\label{troitsk}
\ee

The present spectrometers are unable to improve
the bounds (\ref{mainz}, \ref{troitsk}) substantially.
Further operation
of  Mainz experiment may
allow to reduce the limit down to 2 eV. In this connection a new
experimental project,
KATRIN, is under consideration  with an estimated sensitivity
limit \cite{addressless}
\be
\label{futurelimit}
 m_{\nu_e} \sim  0.3\ \  {\rm eV}.
\ee
In the case of negative  result fromthe  KATRIN searches
one can get after three years of operation  the bound
$m_{\nu_e} \leq 0.35~~ (0.40)$ eV at 90 \%
(95 \%) C.L. \cite{addressless}.

Note that with this bound  KATRIN experiment can explore the range
of neutrino mass which is  relevant
for the Z-burst explanation of the cosmic ray with super-GZK energies
\cite{fargion}.

The aim of this paper is to study the discovery potential of the next
generation tritium beta decay experiments with sensitivity in the
sub-eV range and in
particular,  KATRIN  experiment. We  consider the effects of  neutrino
mass and  mixing  on
the $\beta$-decay spectrum expected for specific neutrino schemes.
We describe the  three-neutrino schemes which are elaborated to explain
the
data on the solar and atmospheric neutrinos as well as  the  four-neutrino
schemes which
accommodate also the LSND result.
We study the bounds that the present
and forthcoming  $2 \beta 0 \nu$-decay searches, as
well as the oscillation experiments can put on  possible tritium decay
results.
We also consider  the implications of future beta decay measurements
for the identification of the neutrino mass spectrum.

The paper is organized as follows.
In section 2  we give a general description of the effect
of massive neutrinos on the beta decay spectrum in the presence of mixing.
In section 3  the  three-neutrino schemes are  explored.
In section 4 we present a general discussion of predictions for the
beta decay in the four-neutrino schemes which explain the LSND
result. We emphasize the importance of the bounds on the beta decay
parameters imposed by Bugey and CHOOZ experiments.
In section 5 we study the properties of the beta decay
in the hierarchical four-neutrino schemes.
In section 6, the non-hierarchical four-neutrino schemes
are considered.  In section 7 we summarize the role that
future beta decay measurements will play in the
reconstruction of the neutrino mass spectrum.
Conclusions are given in section 8.

\renewcommand{\baselinestretch}{1.2}
{\bf \section{Neutrino mixing and beta-decay. The effects of degenerate states}}
\label{meffective}
\renewcommand{\baselinestretch}{2.0}

In presence of mixing,
the electron neutrino is a combination
of
mass eigenstates  $\nu_i$ with masses $m_i$: $\nu_e=\sum_i U_{ei}
\nu_i $. So that, instead of
(\ref{mother}), the
expression for the spectrum is given by
\be
{dN \over dE}=R(E) \sum_i |U_{ei}|^2
[(E_0-E)^2-{m_i}^2]^{1 \over 2} \Theta (E_0-E-m_i),
\label{improve}
\ee
where $ R(E)$ is defined in  (\ref{re}).
The step function, $\Theta(E_0-E-m_i)$,  reflects the fact that a
given neutrino
can be produced  if the available  energy  is larger than its mass.
According to eq. (\ref{improve})
the presence
of mixing leads to distortion of the spectrum which consists of \footnote{In what follows
we will use the terminology elaborated for the ideal Kurie plot without background.}

(a) the kinks  at the electron energy $E_e^{(i)} = E \sim E_0 - m_i$ whose
sizes are determined by $|U_{ei}|^2$;

(b) the shift of the end point to
$E_{ep} = E_0 - m_1$, where $m_1$ is the lightest
mass in the neutrino mass spectrum.
The electron energy spectrum bends at $E \stackrel {<} {\sim} E_{ep}$.

So, in general the effect of mixed massive neutrinos on the spectrum
cannot be described by
just one parameter. In particular, for the three-neutrino scheme, five
independent
parameters are involved: two mixing parameters  and three
masses.

Substantial simplification, however, occurs in the schemes
which explain the solar and atmospheric neutrino data
and have the states with absolute values of masses in the range of sensitivity
(\ref{futurelimit}).  The simplification appears due to  existence
of sets of  quasi-degenerate states.
Indeed, in these schemes
there should be eigenstates with mass squared differences
$\Delta m_\odot ^2 < 2 \cdot 10^{-4}\ \  {\rm eV}^2 $ and $\Delta m_{atm}^2
{\sim} 3 \cdot 10^{-3}~ {\rm eV}^2$.
If the  neutrino masses, $m_i$, are    larger
than 0.3 eV  (\ref{futurelimit}),
the  mass differences
\be
\Delta m \sim \frac{\Delta m^2}{2m }
\ee
turn out to  be smaller than $5 \times 10^{-3}$ eV.
Moreover, $\Delta m/m \sim \Delta m^2/2m^2 \ll 1$, that is,  the
states are strongly degenerate. Since the detectors cannot
resolve such a  small mass split,  different
masses will entail just to one visible kink with
certain effective mass and mixing parameter.
As a  consequence,  the number of relevant parameters
which describe the distortion of the beta spectrum  is reduced to
one or three, depending on the type of the scheme (see sections 3 - 6).

In the Ref. \cite{vissani} it has been shown that  for energies
$E_{\nu_e} \gg m_{\nu_i}$  the distortion of the  electron
energy spectrum in the $\beta$-decay due to non-zero  neutrino mass and mixing
is determined by the  effective mass
\be
m_{eff}={\sum_i m_i |U_{ei}|^2 \over \sum_i |U_{ei}|^2 }.
\label{mix1}
\ee
However, the highest sensitivity to the mass
of $\nu_i$ appears in the energy range  close to the end point where
 $ E_\nu \sim m_i$ and therefore the approximation used in
\cite{vissani} to introduce
$m_{eff}$ does not work.
In what  follows we show that
still it is possible to use the  mass parameter (\ref{mix1}) for  a set of
quasi-degenerate states.

In general,  the neutrino mass spectrum can have   one or more sets
of quasi-degenerate states. Let us consider
one  such a set which contains $n$ states,
${\nu_j}$, $j = i, i+1, ..., i+ n - 1$ with
$\Delta m_{ji} \ll m_j$.  We define $\Delta E$ as the smallest  energy
interval that the spectrometer can resolve. (Note that $\Delta E$
may be smaller than  the  width of
resolution function, and the latter  is about 1 eV in KATRIN
experiment.) We assume that $\Delta m_{ij} \ll \Delta E$.

Let us introduce  the  coupling of this set of the states
with the  electron neutrino as
\be
\label{total}
\rho_e \equiv \sum_j|U_{ej}|^2,
\ee
where $j$  runs over the  states in the set.
We will  show  that the observable effect of such a  set
on the beta spectrum can be described by $\rho_e$ and
the   effective mass
$m_{\beta}$ which can be  introduced in the following way.
Let us consider the interval $\Delta E$ in the region of the highest
sensitivity to the neutrino mass, that is,  the interval of the electron
energies
\be
(E_0 - m_i -\Delta E)  -  (E_0 - m_i)~,
\ee
where $m_i$ is the mass of the lightest state in the set.
The number of events in this interval, $\Delta n$, is given by
the integral
\be
\Delta n = \int_{E_0-m_i-\Delta E}^{E_0-m_i} {dN \over dE} dE.
\label{integ}
\ee
We will define the effective mass $m_{\beta}$ in such a way that
the number of events calculated for the approximate spectrum   with
single mixing parameter $\rho_e$ and mass  $m_{\beta}$, $\Delta n
(\rho_e, m_{\beta})$, reproduces,
with
high precision, the number of events calculated for exact neutrino mass
and mixing spectrum $\Delta n (U_{ej}, m_j) $. That is,
\be
R \equiv {\Delta n (\rho_e, m_{\beta}) - \Delta n (U_{ej}, m_j) \over
\Delta n (\rho_e, m_{\beta})} \ll 1.
\ee
Expanding $\Delta n$  (see Appendix)
in powers  of  $\Delta m_j / \Delta E \ll 1$, where
\be
\Delta m_j \equiv m_j -  m_{\beta}~,
\ee
we obtain:
\be
R \propto
\sum_j |U_{ej}|^2 \frac{\Delta m_{j}}{\Delta E} +
O\left( \left(\frac{\Delta m_{j}}{\Delta E}\right)^2 \right).
\ee
It is easy to see that the first term  vanishes   if
\be
m_{\beta} = {\sum_j m_j |U_{ej}|^2 \over \rho_e}.
\label{m_beta}
\ee
So, for this value of $m_\beta$, $R$ is of the order of $(\Delta
m_{j}/\Delta
E)^2$.
Note that if we
set $m_{\beta}$ to be equal to the value of any  mass from the set or the
average
mass, the difference of the number of events  would  be
of the order of ${\Delta m/\Delta E}$. If  $\Delta E$ is relatively small,
this correction may be significant.
The expression (\ref{m_beta}) is similar to
(\ref{mix1}), but  in (\ref{m_beta}) $j$ runs over a
quasi-degenerate set (not over all the states.) Moreover,   provided
that $\Delta m \ll \Delta E$, the approximation works for all
energies.

In  reality  the background should be taken into account.
However it is easy to see that if the change of the background with
energy in the interval
$\Delta E$ is negligible, our
analysis will be  valid in the presence of the
background, too.

If the  scheme contains  more than one set
of quasi-degenerate states with the
corresponding effective masses $m_{\beta}^q$ and mixing parameters
$\rho_e^q$,
the observable spectrum can be described by   the following expression
\be
{dN  \over dE}=
R(E)\sum_q \rho_e^q \ \ [(E_0-E)^2-{(m_{eff}^q)}^2]^{1 \over 2} \Theta
(E_0-E-{m_{eff}^q}),
\ee
where $q$  runs over the  sets.
Each set of  quasi-degenerate states will produce a
single kink at the electron energy  $E^q \sim E_0 - m_{\beta}^q$ with  the
size of the kink determined by $\rho_e^q$.
The set with the lightest masses
leads to bending of spectrum and the shift of the end point.

\section{Three neutrino scheme}

Let us consider the three-neutrino schemes which explain the solar
and atmospheric neutrino results.
In the case of mass hierarchy, $m_1 \ll m_2 \ll m_3 $,
the largest mass,
$m_3 \simeq \sqrt{\Delta m_{atm}^2}=(4-7) \times 10^{-2}$ eV,
is too small to produce  any observable effect in the planned tritium
decay  experiments (see eq. (\ref{futurelimit})).

If $m_3$ is in the sensitivity range of   KATRIN
experiment ($m_3 \geq$ 0.3 eV), the mass spectrum
should be quasi-degenerate. Indeed,
$$
{\Delta m_{31} \over m_3} \simeq {\Delta m_{atm}^2 \over 2 m_3^2} \leq 0.03.
$$
Moreover, from the unitarity condition we get  the coupling parameter
\be
\rho_e = \sum_{j = 1,2,3} |U_{ej}|^2 = 1.
\ee
Therefore  the effect of non-zero neutrino  masses and mixing  on the
$\beta$-decay
spectrum is  characterized  by  unique parameter - the effective mass
\be
m_{\beta} =   \sum_{j = 1,2,3} m_j |U_{ej}|^2 \simeq m_3.
\ee
Correspondingly,  the distortion of the $\beta$-decay spectrum  consists
of a
bending of the spectrum and shift of the end point determined
by $m_{\beta}$ ($E_0 \rightarrow E_0 - m_{\beta}$),
as in the case of $\nu_e$  with definite mass and without mixing.
Let us consider the bounds on
$m_{\beta}$ imposed by the $2\beta 0 \nu$-decay
searches
and the oscillation experiments. (The $2\beta 0 \nu$-decay in schemes
with three degenerate neutrinos has been extensively discussed before
\cite{petcov}, \cite{3nu}).
Assuming that neutrinos are  Majorana particles we get from (\ref{mee})
and (\ref{m_beta}) the relation between
the effective masses in the beta decay and the double beta decay:
\be
m_{ee} \simeq m_{\beta}||U_{e1}|^2 + e^{i\phi_2} |U_{e2}|^2 +
e^{i\phi_3}|U_{e3}|^2|,
\label{phaseless}
\ee
where $\phi_2$ and $\phi_3$ are the relative CP-violating phases
of the contributions
from the second  and the  third mass eigenstates.

According to the CHOOZ bound \cite{Apollo}  which is confirmed by
the slightly weaker bound obtained in Palo Verde experiment \cite{Boehm},
one of the squared mixing elements (let us take
$|U_{e3}|^2$ for definiteness) must be smaller than 0.05.
The  other two elements are
basically determined by the mixing angle $\theta_{\odot}$ responsible for
the solution of the solar neutrino problem, so
that the eq. (\ref{phaseless})
can be rewritten as
\be
m_{ee} = m_{\beta} \left| (1-|U_{e3}|^2) (\cos ^2 \theta_{\odot} +
e^{i\phi_2} \sin ^2 \theta_{\odot}) +   e^{i\phi_3} |U_{e3}|^2 \right|.
\label{77}
\ee
{}From this equation we find the following bounds  on the beta decay mass
(see
also \cite{onethird}):
\be
m_{ee} < m_{\beta} <
{m_{ee} \over
\left| |\cos 2 \theta_\odot|(1-|U_{e3}|^2) - |U_{e3}|^2 \right| } ,
\label{bound}
\ee
where the upper bound corresponds to the maximal cancellation of
the  different terms in (\ref{77}).

\begin{figure}
\begin{center}
\vskip -3.2cm
\hskip -7.0cm
\parbox[c]{3.5in}
{\mbox{
\qquad\epsfig
{file=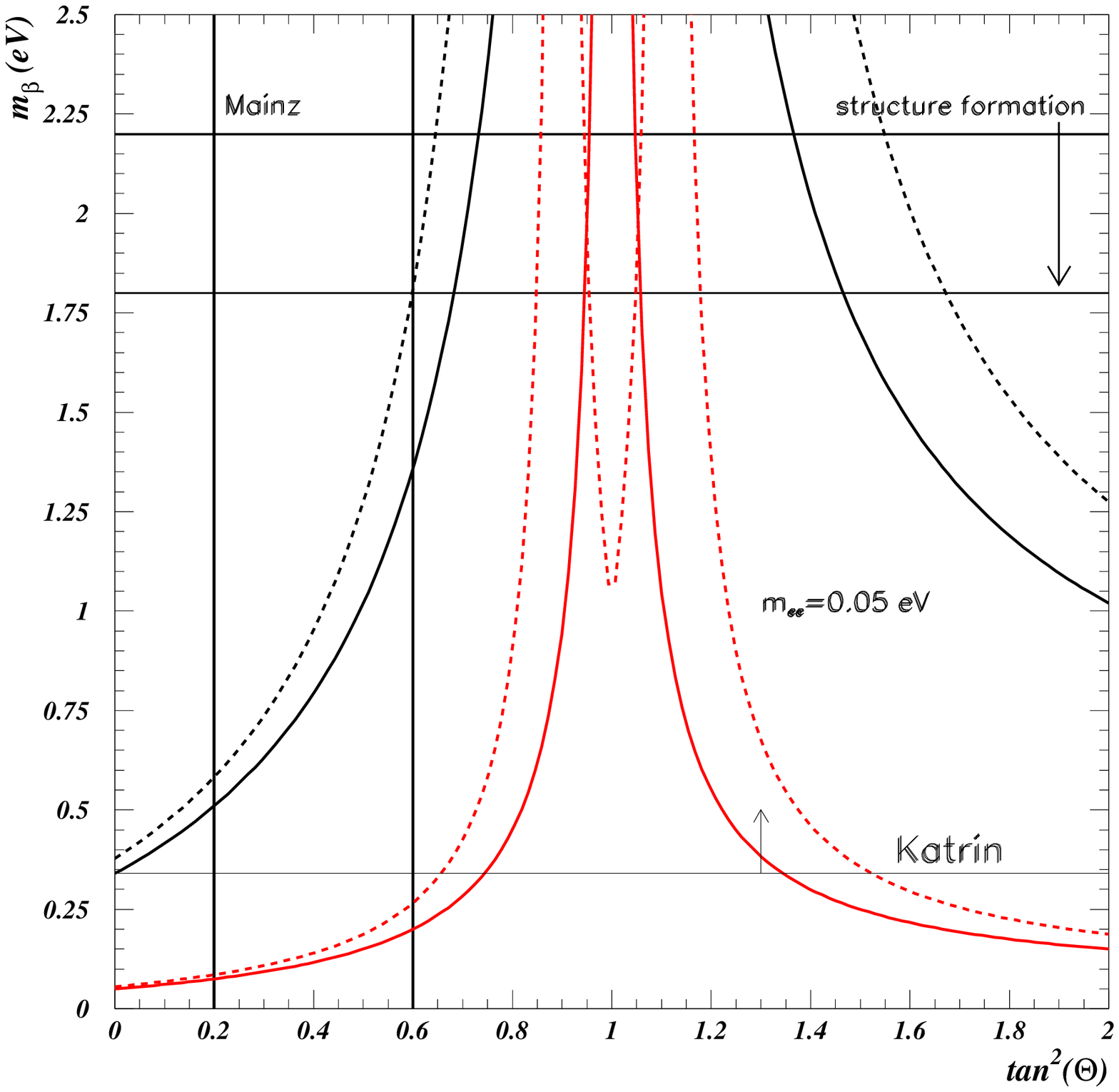,width=1.6\linewidth,height=1.6\linewidth}}}
\end{center}
\vskip -0.8 cm
\caption{ The bounds on the effective $\beta$-decay mass, $m_{\beta}$,
in the
$3\nu$-scheme with
mass degeneracy.
Shown are the  upper bounds from the $2 \beta 0 \nu $-decay
as the
functions
of mixing angle relevant for the solution of the solar neutrino problem (see
eq. (\ref{bound})).
  The upper solid (dashed) line corresponds to the present bound $m_{ee}
\leq 0.34$ eV and $|U_{e3}|^2=0$
($|U_{e3}|^2$=0.05). The  lower solid  (dashed) line corresponds to
$m_{ee} \leq  0.05$~eV  and $|U_{e3}|^2=0$ ($|U_{e3}|^2$=0.05).
These lines are drawn in assumption of strong degeneracy of neutrino
masses.
The vertical lines mark the 90 $\%$ C.L.
borders of the LMA solution region. Shown also are the present upper
bound on the neutrino mass from structure formation  \cite{Liu} and the
sensitivity limit of KATRIN experiment.  } \label{xfig1}
\end{figure}

The bounds  (\ref{bound}) are shown in fig. \ref{xfig1}.
(See  also \cite{petcov}.)
The following comments are in order:

1) For zero value of $U_{e3}$ the  weakest bound on
$m_{\beta}$ from the double beta decay  appears at maximal mixing:
$\tan^2 \theta_{\odot} = 1$. For non-zero $U_{e3}$  the points of the
weakest bound shift to $\tan^2 \theta_\odot \simeq 1 \pm 2 |U_{e3}|^2   $.
In the
vicinity of these points,
the upper bound on $m_{\beta}$  is given by the present beta decay
result (see eq. (\ref{mainz})).

2) Taking  the best fit values of $\theta_\odot$ from
the various large mixing solutions of the  solar
neutrino problem  \cite{gzz} we find from eq. (\ref{bound}) the following
bounds:
\ba
m_{\beta} < \left\{ \matrix{ 0.67-0.74 & {\rm eV} & {\rm LMA} \cr
1.6-2.2 & {\rm eV} & {\rm LOW} \cr
1.0-1.3 & {\rm eV} & {\rm VAC}  \cr } \right. ,
\ea
where we have used  $m_{ee} \leq $0.34 eV (\ref{dbbound}) and the two
numbers in each line
correspond to $|U_{e3}|^2$= 0 and  0.05, respectively.
Note that already existing data on the $2 \beta 0 \nu$-decay give  bounds
(at the best fit points)
which are stronger than the present bound from direct measurement.

Moreover, for $m_{ee} < 0.07$ eV which can be achieved already by CUORE
 experiment \cite{cuore}, the bound on $m_{\beta}$ from
$2 \beta 0 \nu$-decay in the LMA
preferable region of $\tan^2 \theta$ is below the sensitivity  of
KATRIN experiment. Therefore, the positive  result of  KATRIN
experiment
(and identification of  the LMA solution of the solar neutrino problem)
will lead to exclusion of such  a $3\nu$-scheme.

3) For the  SMA solution of the  solar neutrino problem we get
$m_{\beta}  \simeq m_{ee}$, and consequently,  according to
the bound (\ref{dbbound}):
$m_{\beta} \leq 0.34$ eV. So, the expected range of $m_{\beta}$  only
marginally overlaps with the KATRIN sensitivity region.
Thus, if the SMA solution is identified and the LSND result is not
confirmed,
favoring the three neutrino scheme, the chance for
observation of  the beta spectrum distortion in KATRIN experiment is
rather small.

4) A positive signal in the $2 \beta 0 \nu$-decay searches will have
important implications for the tritium decay measurements:

a).  According to (\ref{bound}),  it gives a lower bound on $m_{\beta}$
independently  of the
solution of the solar neutrino problem: $m_{ee} \leq m_\beta $.

b). If  the values of $m_{ee}$, $m_{\beta}$ and $|U_{e3}|^2$ are
measured, we will be able to  determine  the CP-violating phase $\phi_2$
in (\ref{77}):
$$
\sin ^2 \frac{\phi_2}{2} =
\frac{1}{\sin^2 2 \theta_\odot}
\left[
1 - \left({m_{ee} \over m_{\beta}}\right)^2 - 2 |U_{e3}|^2
\left(\left({m_{ee} \over m_{\beta}}\right)^2
\pm {m_{ee} \over m_{\beta}}\right)
\right]
$$
$$
\simeq
\frac{1}{\sin^2 2 \theta_\odot}
\left[1- \left({m_{ee} \over m_{\beta}}\right)^2 \right] ,
$$
where ($\pm$) sign of the last term reflects an uncertainty  due
to the phase of $U_{e3}^2$  ($\phi_3$ in (\ref{phaseless})).

c). If $m_{\beta}$ turns out to be smaller  than
$m_{ee}$, we will  conclude that there are some additional
contributions to the   $2\beta 0 \nu$-decay unrelated to the Majorana
neutrino mass.

If future $\beta$ decay measurements  with sensitivity
(\ref{futurelimit}) give a negative result, the largest part of the
allowed mass range of the 3$\nu$-scheme   with {\it strong}
degeneracy will be excluded. Still a small interval
($m_\nu \sim 0.1-0.3$ eV) will be uncovered.
This will have  important implications for the theory of the neutrino
masses.

In fig. \ref{xfig1} we show also the upper
bound on $m_\beta$ from data on the
large scale structure of the Universe obtained in \cite{Liu} for
values of total matter contribution to the energy density of the Universe,
$\Omega_m=0.4$ and the reduced Hubble  constant, $h=0.8$.

Without $\beta$-decay measurements the absolute value
of the neutrino mass can be
determined and the scheme can be identified
provided that all the following conditions are  satisfied:

-  no effect of sterile neutrinos is observed,

-  the SMA solution is established as the solution
   of the solar neutrino problem,

-  the neutrinoless double beta decay  gives a
positive result with
   $m_{ee}$ close to the present upper bound.

In this case $m_\nu=m_{ee}$.
However, not too much room is left for such a possibility keeping
in mind that recent solar neutrino data disfavor the   SMA solution.
For all large mixing solutions of the solar neutrino problem,
$m_{ee}$ gives only the lower bound on the absolute scale of masses.

Let us stress that,\  even in the case of the SMA solution
one should make the assumption that  there are no  additional
contributions
to the $2 \beta 0 \nu$-decay apart from the exchange of  light Majorana
neutrinos.
In this scheme there are  no ``test equalities",
that is, the relations between   $m_{ee}$
and the oscillation parameters, $\Delta m^2$, $\theta$,  which could allow
to check this assumption independently.
Furthermore, the determination of $m_{ee}$ and therefore $m_\nu$ will be
restricted by uncertainties in the nuclear matrix elements. Thus,
the study  of the
beta spectrum is the only way to measure the absolute scale of neutrino mass
without ambiguity.

Clearly, if the LSND result is confirmed the scheme will be excluded.

\section{4-$\nu$ schemes:  Bugey, CHOOZ and LSND bounds}

\begin{figure}
\parbox[c]{3.5in}
{\mbox{\qquad\epsfig{file=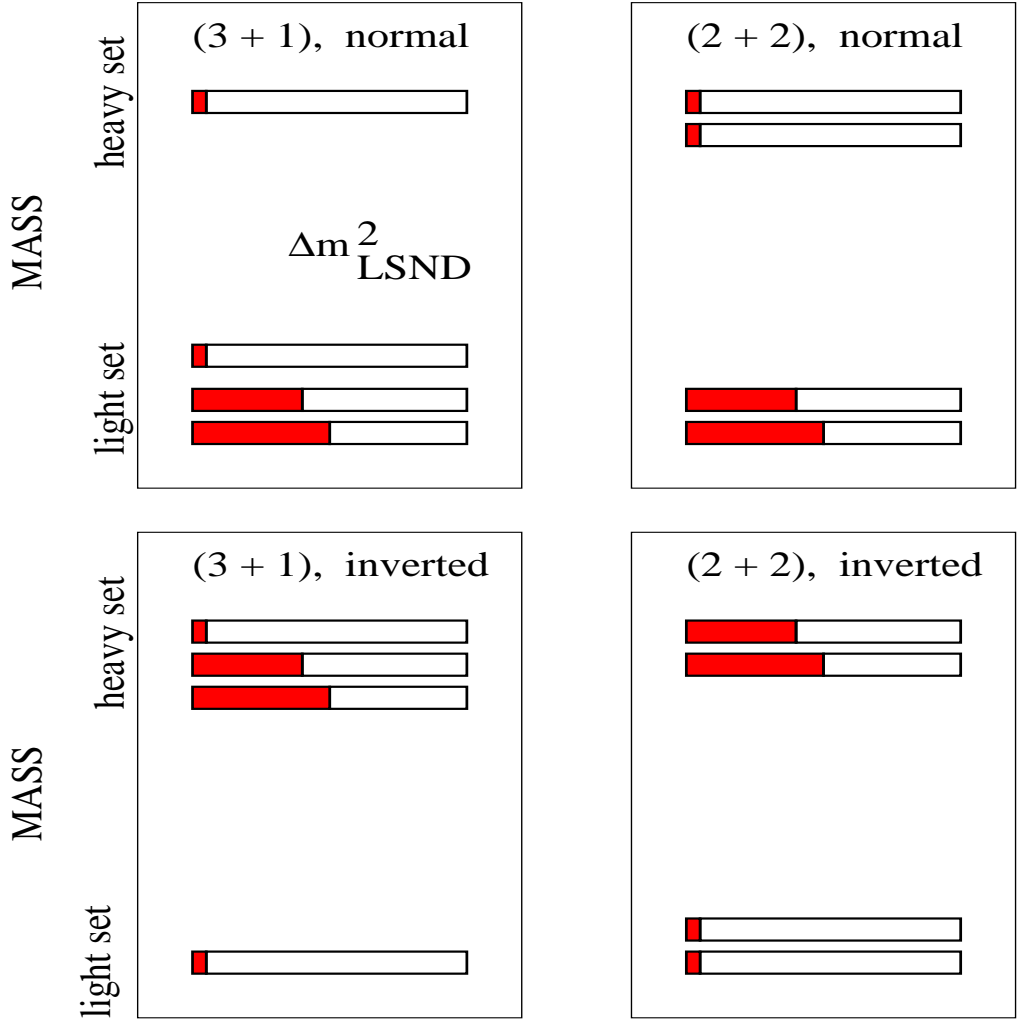,width=1.6\linewidth,height=1.6\linewidth}}}
\caption{The
four neutrino schemes of  mass and mixing.
The boxes correspond to the  mass eigenstates.
The position of the box  in the vertical axis determines the mass.
The shadowed parts  of  the boxes indicate the
amount  of the electron flavor in the corresponding
eigenstate $\nu_i$, that is, $|U_{ei}|^2$. The solar pair is formed by
the two strongly degenerate
states with $\Delta m^2_\odot$ and significant amount of the electron
flavor.
In the hierarchical schemes the effective mass  of the light
set is much smaller than the mass of the heavy set.
In the non-hierarchical schemes these two masses are
comparable. For definiteness we show distribution of the electron flavor in
the solar pair which corresponds to the large mixing solution of the solar
neutrino problem.} \label{box}
\end{figure}

 Four-neutrino schemes, which explain  the LSND result
in terms of oscillations, have  two sets of mass eigenstates
separated by $\Delta m_{LSND}^2$ (see fig. \ref{box}). Hereafter,
we call them the  light set of states
and the heavy  set of states.
Let us consider the heavy set. The
masses  of states in this set  are equal or larger  than
$\sqrt{\Delta m_{LSND}^2}$. The
mass differences
 are equal or smaller than
$${\Delta m_{atm}^2 \over  2\sqrt{\Delta m_{LSND}^2}}
\ \ {\rm or/and}\ \    {\Delta m_{\odot}^2 \over 2 \sqrt{\Delta
m_{LSND}^2}}.$$
Both splits are much smaller
than   the energy resolution $\Delta E$ as well as  masses themselves.
So,  the states in the heavy  set are
quasi-degenerate and their effect on the beta spectrum
can be characterized  by $m_{\beta}^h$ and $\rho_e^h$ given in
eqs. (\ref{total})  and (\ref{m_beta}).

In the (2 + 2) schemes both the heavy and light
sets contain two states,
whereas in the
(3 + 1) scheme
one set contains 3 states, while the other set consists of  only
one state (see fig. 2).

The $\nu_e$ oscillation disappearance experiments, Bugey \cite{bugey}
and CHOOZ \cite{Apollo},
 impose a direct and very strong
bound on $\rho_e^h$, and therefore on the expected effects
in $\beta$-decay in all 4$\nu$-schemes.
Since Bugey and CHOOZ experiment do not resolve small mass
squared differences,
$\Delta m_{atm}^2$ and $\Delta m_\odot ^2$,
their results can be described by  2$\nu$-oscillations
with a unique mass squared difference
$\Delta m^2 \simeq  \Delta m^2_{LSND}$ and the
effective mixing parameter
$$\sin^2 2\theta_{eff}=4
\sum_i
|U_{ei}|^2(1-\sum_i |U_{ei}|^2),$$
where the sum runs over the heavy (or light) set.
Using the definition of  $\rho_e^h$ in eq. (\ref{total}) we can
rewrite the mixing parameter as
\be
\sin^2 2 \theta_{eff} = 4
\rho_e^h (1-\rho_e^h). \label{buchooz}
\ee
Thus,  the negative results of the oscillation searches in  Bugey and
CHOOZ experiments
give immediate bound on  $\rho_e^h$ as a function of $\Delta m^2_{LSND}$
(see figs. \ref{xfig2} - \ref{xfig6} ).

For the range of masses relevant for
LSND experiment ($\Delta m^2 > 0.2$ eV$^2$)  two possibilities follow
from
(\ref{buchooz})
and the Bugey or CHOOZ bounds:

1). Small $\rho_e^h$:
\be\rho_e^h < 0.027.
\label{norm}
\ee
That is, the  admixture of $\nu_e$ in the heavy  set is very small
and the electron flavor is distributed mainly in the light  set.
This corresponds to the schemes with normal mass hierarchy
(or to normal ordering of states
in the non-hierarchical schemes). Let us recall that in the schemes with
normal hierarchy (order of states) the light set contains the pair of states
which  are separated by $\Delta m_\odot$. This pair is responsible for the
conversion
of the solar neutrinos, and for brevity, we will call it  "solar
pair".
According to (\ref{norm}),  in
this class of schemes one expects  small kink
at $E_e \sim  E_0 - m_{\beta}$, where
$m_{\beta} \geq \sqrt{\Delta m_{LSND}^2}$.
Here, the inequality corresponds to the
non-hierarchical case (see sect. 6). Also in the case
of non-hierarchical scheme one
predicts an observable shift of the end point associated
to the masses of the light set.

2). Large $\rho_e^h$:
\be
1 - \rho_e^h < 0.027.
\label{inv}
\ee
The admixture of  $\nu_e$ in the heavy set is close to one: the electron
flavor is mainly distributed in the heavy set. This corresponds to
the schemes with
inverted mass hierarchy or inverted ordering of states.
The effect in the beta spectrum consists of a ``large" kink
with size close to 1 at  $E_e \sim E_0 - m_{\beta}$.
Above $E_e \sim E_0 - m_{\beta}$
the spectrum  has a tail related to emission of the
neutrinos from the light set. The rate of events in the tail
is suppressed by factor given in eq. (\ref{inv}).
If the  tail is
unobservable,
the whole effect will look like the effect of the electron neutrino with a
unique
definite mass $m_{\beta}$. (And it is similar to the effect in the scheme
with three degenerate neutrinos.)

We will further discuss these possibilities in the
sections 5-7.\\
Let us consider  implications of the LSND result itself for the $\beta$-decay
searches.
Apart from providing the mass scale in the range of sensitivity of future
$\beta$-decay experiment, it imposes an important bound
on the relevant mixing parameters.

In the  4$\nu$-schemes under consideration the oscillations
in LSND experiment are reduced to two neutrino oscillations with
$\Delta m^2 \simeq  \Delta m^2_{LSND}$ and the effective mixing
parameter
\be
\sin ^2 2 \theta_{LSND} = 4 |\sum_{j \in h} U_{\mu j} U_{e j}^*|^2 ~=
4 |\sum_{j \in l} U_{\mu j} U_{e j}^*|^2 ,
\ee
where summations  run over the states of the heavy set in the first
equality
and of the light  set in the second equality.
Using Schwartz inequality we get
\be
\sin^2 2 \theta_{LSND} \leq 4 (\sum_{i \in h} |U_{ei}|^2)(\sum_{j \in h}
|U_{\mu j}|^2)  = 4 \rho_e^h \rho_{\mu}^h,
\label{ineq_sch}
\ee
or equivalently,
\be
\sin^2 2 \theta_{LSND} \leq 4 (\sum_{i \in l} |U_{ei}|^2)(\sum_{j \in l}
|U_{\mu j}|^2)  = 4 (1-\rho_e^h)(1- \rho_{\mu}^h),
\label{addd}
\ee
where
\be
\rho_\mu^h \equiv \sum_{j \in h} |U_{\mu j}|^2 \label{rhomu}
\ee
is the coupling of the heavy  set  with the muon neutrino.
The upper bound on  $\rho_\mu^h$ ( or upper bound on $1-\rho_\mu^h$,
depending on the scheme)  follows from
CDHS experiment  at high $\Delta m^2$ \cite{cdhs},
and from the atmospheric neutrino studies at
low $\Delta m^2$  \cite{kajita}.
{}From eq. (\ref{ineq_sch}) we get a  lower bound on $\rho_e^h$:
\be
\rho_e^h  > \frac{\sin^2 2 \theta_{LSND}}{4 \rho_{\mu}^h},
\label{rho_bound}
\ee
where both $\sin^2 2 \theta_{LSND}$ and $\rho_{\mu}^h$
are  functions of $\Delta m^2$.
Implications of this bound for the $\beta$-decay measurements
strongly depend on  specific scheme,
and we will discuss them in sections 5 and 6.

The  character of  the  distortion of the spectrum and
the sizes of the effects depend on

1)  the structure of   the scheme: hierarchical or non-hierarchical;

2)  the type of hierarchy (ordering of levels): normal or inverted;

3) the number of states in the heavy and light sets.\newline
In what follows we  consider  possible 4$\nu$-schemes in order.

\section{Four-neutrino schemes with mass hierarchy}

In the hierarchical schemes, the masses of the states from  the  light  set
are much smaller than the sensitivity limit 0.3 eV
(\ref{futurelimit}).  In the (2 + 2) schemes they
are restricted by the atmospheric ($m^l \leq \sqrt{\Delta m_{atm}^2}$) or
solar ($m^l \leq \sqrt{\Delta m_{\odot}^2}$) neutrino mass  scales.
Therefore   the observable distortion of the spectrum is only due to
 effect of the heavy set with the effective mass
\be
m_{\beta} \approx \sqrt{\Delta m_{LSND}^2}.
\ee

As we have mentioned in the previous  section,
the character of distortion of the beta decay
spectrum depends, first of all,  on the type of hierarchy.

\subsection{Schemes with normal mass hierarchy}

\begin{figure}
\begin{center}\hskip -7.0cm
\parbox[c]{3.5in}{\mbox{\qquad\epsfig{file=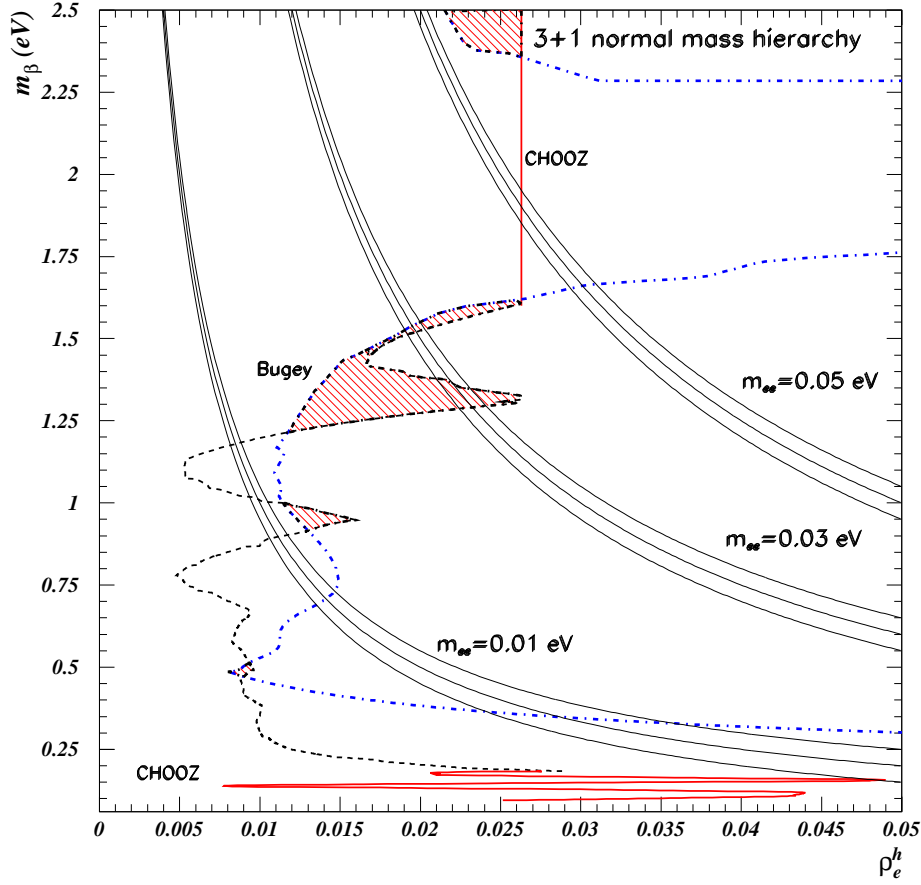,width=1.5\linewidth,height=1.5\linewidth}}}
\end{center} 
\vskip -1.0cm
\caption{
The bounds on $m_{\beta}$ and
$\rho_e^h$ in the (3+1) scheme with normal mass
hierarchy.
The dashed curve and solid lines attached to it (from below and above) show the
upper bound on $\rho_e^h$ from
Bugey and CHOOZ experiments, respectively.
The LSND lower bound (see eq. (\ref{rho_bound}))
is shown by dot-dashed
curves. The allowed regions are shadowed.
  The triplets of  solid lines show the upper bounds on
$m_\beta$ assuming that
 future $2 \beta 0 \nu$-decay searches will give
$m_{ee} \leq$ 0.01, 0.03 and 0.05 eV. The central
line in each triplet corresponds to the
contribution from the heaviest mass eigenstate
(eq. (\ref{mee4})) and the other two lines show the
uncertainty due to the contribution of light
states in the modification of the scheme in which the mass hierarchy of the three
light states is inverted (see eq. (\ref{ni})).}
\label{xfig2}
\end{figure}

\begin{figure}
\begin{center}
\vskip -3.0cm
\hskip -7.0cm
\parbox[c]{3.5in}
{\mbox{\qquad\epsfig{file=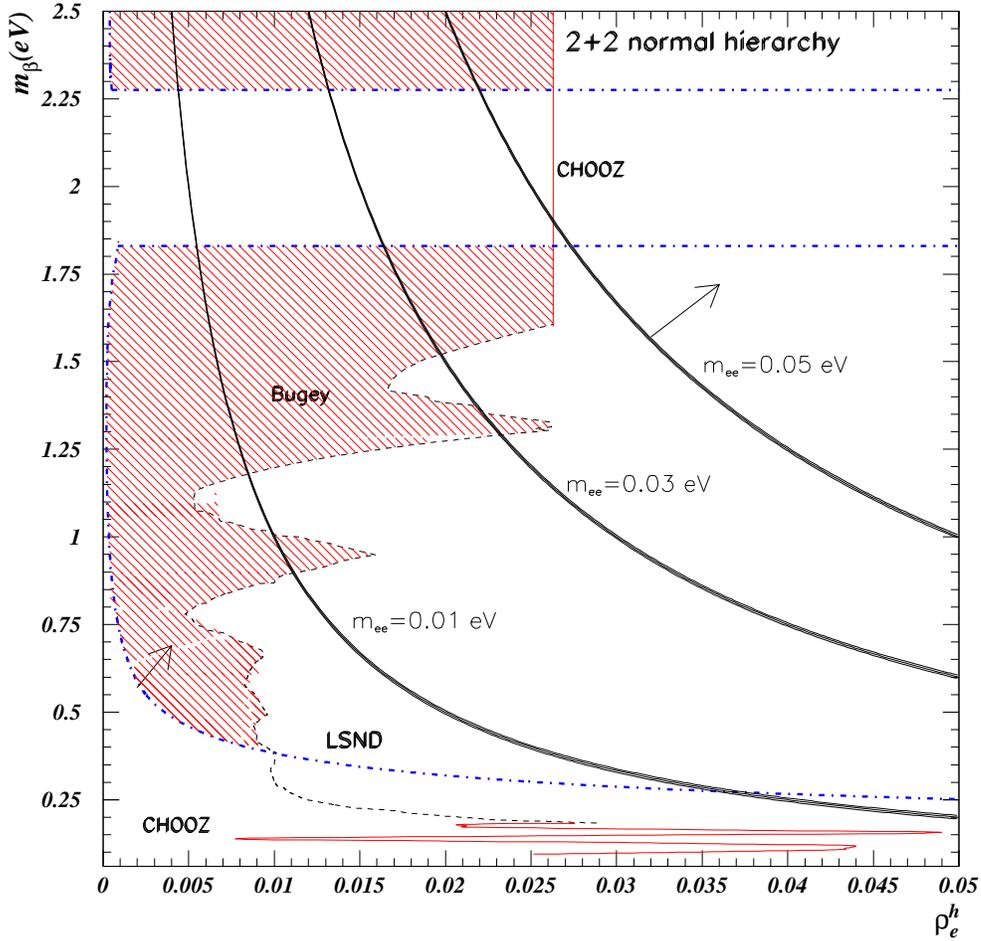,width=1.6\linewidth,height=1.6\linewidth}}}
\end{center}
\vskip -0.8cm
\caption{The bounds on $m_{\beta}$ and
$\rho_e^h$ in the (2+2) scheme with normal mass
hierarchy. The dashed curve and solid lines attached to it  show the
upper bounds on $\rho_e^h$ from
Bugey and CHOOZ experiments, respectively.
The LSND lower bound (see eq. (\ref{rho_bound})) is shown by dash-dotted line.
The allowed regions of parameters are shadowed.
The triplets of solid lines show the lower bounds on $m_\beta$ from
a positive signal
in future  $2 \beta 0 \nu $-decay searches which would correspond to
$m_{ee}$=0.01, 0.03 and
0.05 eV. The central lines correspond to contribution to $m_{ee}$ from the heavy set
only.}
\label{xfig3}
\end{figure}

In the  schemes with the normal hierarchy
(both in the (2 + 2) and (3 + 1) cases)
the electron flavor is in the light set   and  $\rho_e^h$
is strongly
restricted by  the Bugey result  (see eq. (\ref{norm}) and  figs.
\ref{xfig2}, \ref{xfig3}).
The beta decay spectrum  has  only a ``small" kink with $\rho_e^h < 0.027$
at $E = E_0- \sqrt{\Delta m_{LSND}^2}$.

Additional restrictions on the $\beta $-decay parameters may appear
depending on whether the scheme is of the (2 + 2) or (3 + 1) type.\\

{\it 1). The (3 + 1) scheme.}
In  this scheme   one gets a  substantial lower bound
on  $\rho_e^h$ from the  LSND result (see eqs. (\ref{ineq_sch}),
(\ref{rho_bound})). Indeed, in this scheme
$\rho_\mu^h$ is restricted from above by the CDHS result \cite{cdhs}.
Inserting the bound on $\rho_\mu ^h$
into eq. (\ref{rho_bound}) we get  the lower bound on $\rho_e^h$  which
is  close to the upper Bugey bound or, for certain ranges of
$\sqrt{\Delta m_{LSND}^2}$, even above it. So that
only
certain ranges of $m_\beta$ are allowed (see fig.
\ref{xfig2}).
This is a manifestation of the fact that in
the (3 + 1) scheme an explanation of the LSND result requires the $\nu_e$
admixture in the isolated state to be  at the level of the upper Bugey bound
\cite{chance}.

Let us consider  implications of the $2 \beta 0 \nu$-decay search.
The contribution to $m_{ee}$ from the fourth  (isolated) state
dominates. It  can be estimated as:
\be
 m_{ee}^{(4)}  = \sqrt{\Delta m_{LSND}^2}  |U_{e4}|^2
\sim  (0.005 \ \ -\ \   0.05)~ {\rm eV}~.
\label{test2}
\ee

In the  hierarchical case  with
$m_2=\sqrt{\Delta m_\odot^2}$,  the contributions
from other mass eigenstates can be estimated as
$m_{ee}^{(3)} = \sqrt{\Delta m_{atm}^2} |U_{e3}|^2 < 3.5 \cdot 10^{-3}$ eV
and $m_{ee}^{(2)} \approx \sqrt{\Delta m_\odot^2} \sin^2 \theta_\odot < 7
\cdot 10^{-3}$ eV.
Hence
\be
m_{ee} \approx m_{ee}^{(4)}  =  m_{\beta} \rho_e^h~, \label{mee4}
\ee
or $m_\beta=m_{ee} / \rho_e^h$.

A version of the scheme is possible in which the
mass hierarchy in the light set is inverted,
so that the states which contain the electron
flavor have masses
$m_2 \simeq m_3 \simeq \sqrt{\Delta  m_{atm} ^2}$  and $m_1 \ll m_2$.
In this case we have
\be
m_{ee}\simeq \left| m_4 |U_{e4}|^2 e^{i\delta} +
\sqrt{\Delta m_{atm} ^2}(\cos ^2 \theta_\odot
e^{i\alpha} + \sin ^2 \theta_\odot) \right|,
\label{ni}
\ee
and the contribution from the light set can be  comparable to
the contribution from the 4th state.
The corresponding lines are shown in fig. \ref{xfig2}. According to the
figure  one expects $m_{ee}$ to be
substantially below the present bound: We find
$m_{ee} \sim 0.005$ eV, $\sim 0.015$ eV,
$\sim 0.015 \div 0.03$ eV and $\sim 0.06$ eV for the allowed
``islands"  of $m_\beta$ and $\rho_e ^h$ (from smallest to largest
$m_\beta$). Clearly, the  observation of
$m_{ee}$ near its present experimental bound will exclude the scheme.

{}For the  SMA  solution of the solar neutrino  problem, eq. (\ref{ni})
reduces to
$$
m_{ee}\simeq \left| \sqrt{\Delta m_{LSND} ^2} \rho_e^h e^{i \delta}  +
\sqrt{\Delta m_{atm} ^2} \right|.
$$
In principle, using this equation one can determine the
relative phase $\delta$.

For large angle  solutions of the  solar neutrino problem
still significant  contributions can come from the solar pair of states
and therefore two different phases ($\delta$ and $\alpha$) are involved in the
determination of
$m_{ee}$ (see (\ref{ni})).\\

{\it 2). The (2 + 2) scheme.}
In the  (2 + 2) scheme with normal mass  hierarchy  the mass
difference of the heavy set is given by $\Delta m_{atm}^2$,
 and
$\nu_\mu$ is
distributed, mostly, in the heavy set. According to the CHDS bound,
$\rho_\mu^h$
should
be  close to 1  in this scheme. Hence,
  the LSND bound is less restrictive:
\be
\rho_e^h > \frac{1}{4} \sin^2 2\theta_{LSND},
\ee
and
$\rho_e^h$ can be as small as $(2 - 3) \cdot 10^{-4}$ (see fig.
\ref{xfig3}). The effect of the
kink is unobservable for such a small $\rho_e^h$.

Let us consider bounds from  the $2 \beta 0 \nu$-decay searches.
The Majorana mass of the electron neutrino can be written in the following form
\be
m_{ee} \simeq m_{\beta} \left|
(U_{e3}^2 + U_{e4}^2) +
\frac{\sqrt{\Delta m_\odot ^2}}{m_{\beta}}
\sin^2 \theta_\odot (1 - |U_{e3}|^2 - |U_{e4}|^2)
\right|.
\label{41}
\ee
Neglecting the contribution of light neutrinos
($\sqrt{\Delta m_\odot ^2}  \sin^2 \theta_\odot<7 \cdot  10^{-3}$
eV),
we obtain from (\ref{41}) the following bound on the effective
$\beta$-decay  mass:
\be
{m_{ee} \over \rho_e^h} < m_{\beta}  <
{m_{ee} \over | |U_{e3}|^2 - |U_{e4}|^2|}.
\label{2beta_22}
\ee
If the
neutrinoless double beta decay  is discovered,
this inequality will put a strong lower bound on
$m_{\beta}$:

\be
m_{\beta} > 25 m_{ee},
\ee
where we have used the bound on $\rho_e^h$ from the Bugey experiment.
For $m_{ee} = 0.1$ eV we get $m_{\beta} = 2.5$ eV which is in
the upper  allowed region of the LSND experiment. For this
reason,
one cannot expect $m_{ee} > 0.1$ eV in this scheme.

\subsection{Schemes with inverted  mass hierarchy}

In the schemes with inverted hierarchy, the
electron flavor is distributed mainly in the heavy set
(see fig. \ref{box}). Therefore
the $\beta$- decay spectrum should have a
 ``large" kink with parameters $\rho_e^h \simeq 1$ and
$m_{\beta}^h \simeq \sqrt{\Delta m_{LSND}^2}$.
The light  set  leads to appearance
of the tail at $E_e > E_0 - m_{\beta}^h$ with the suppressed
rate determined by  $(1- \rho_e^h)  < 0.027$.
To detect the signal from the tail one may
search for the integral
effect above $E_0 - m_4$.

\begin{figure}
\begin{center}\hskip -7.0cm
\parbox[c]{3.5in}
{\mbox{\qquad\epsfig{file=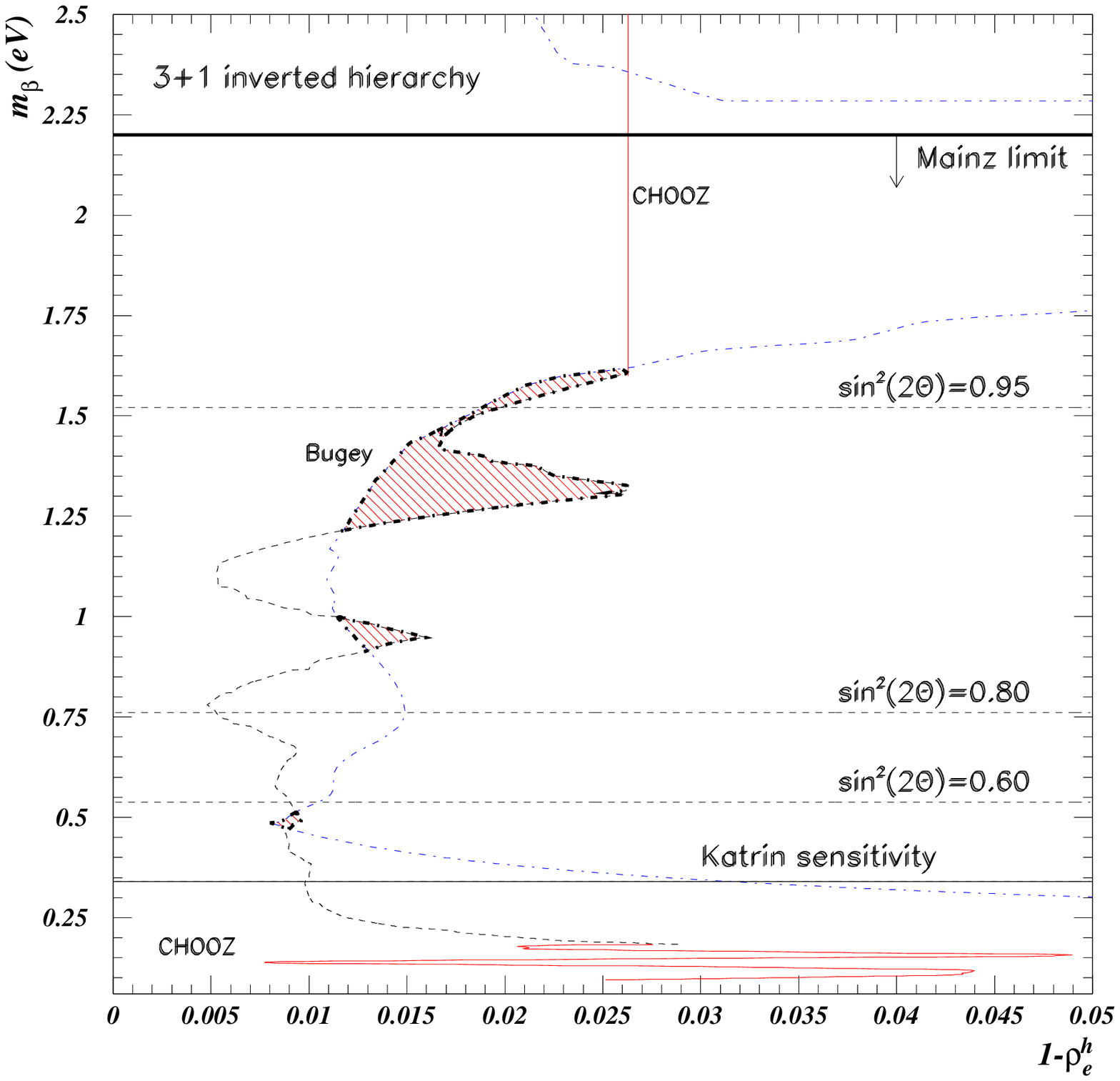,width=1.5\linewidth,height=1.5\linewidth}}}
\end{center}
\caption{
The bounds on $m_{\beta}$ and
$\rho_e^h$ in the (3+1) scheme with inverted mass
hierarchy. The dashed curve and solid lines attached to it
(from below and above)
show the
upper bound on $\rho_e^h$ deduced by
Bugey and CHOOZ experiments, respectively.
The LSND lower bound (see eq. (\ref{rho_bound}))
is shown by dot-dashed lines.
The allowed regions are  shadowed.
The horizontal dashed lines are the upper bounds
on $m_\beta$  from the $2 \beta 0 \nu$-decay
searches which correspond to $m_{ee}<0.34$ eV, $U_{e3}=0$,
and different values of
sin$^2 2\theta_\odot$.}
\label{xfig4}
\end{figure}

\begin{figure}
\begin{center}
\vskip -3.0cm
\hskip -7.0cm
\parbox[c]{3.5in}
{\mbox{\qquad\epsfig{file=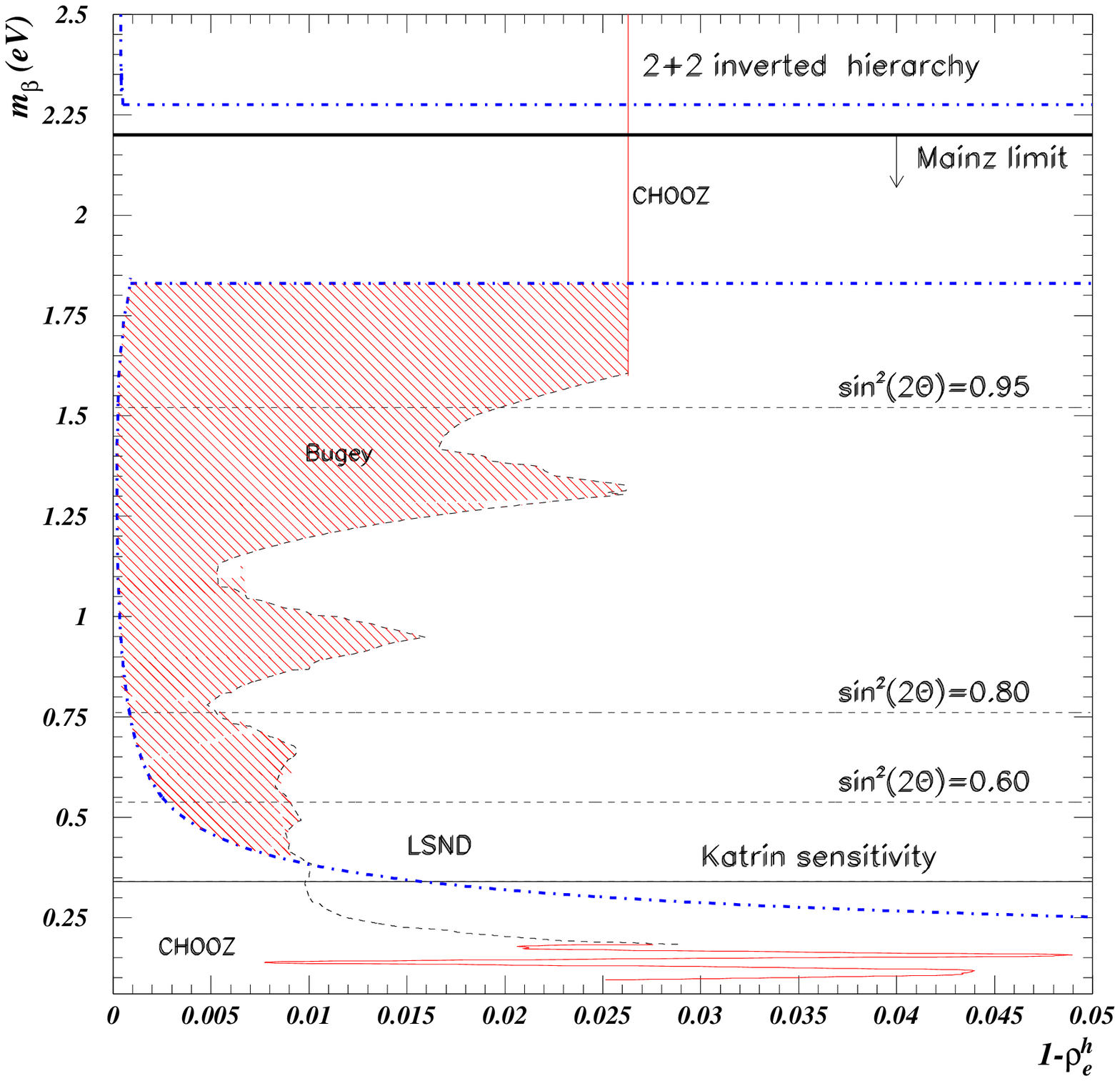,width=1.7\linewidth,height=1.6\linewidth}}}
\end{center}
\vskip -0.8cm
\caption{
The bounds on $m_{\beta}$ and
$\rho_e^h$ in the (2+2) scheme with inverted mass
hierarchy. The dashed curve and solid lines attached to it (from below and above)
show the
upper bound on $\rho_e^h$ deduced by
Bugey and CHOOZ experiments, respectively.
The LSND lower bound (see eq. (\ref{rho_bound})) is shown by dot-dashed lines.
The allowed regions are  shadowed.
The horizontal dashed lines are the upper bounds
on $m_\beta$  from the $2 \beta 0 \nu$
searches which correspond to $m_{ee}<0.34$ eV
and different values of
sin$^2 2\theta_\odot$. We assume  that solar pair of states  gives the
dominant
contribution (see eq. (\ref{U=0})).  The curve for the SMA solution (not
shown here) practically coincides with the curve of KATRIN sensitivity.}
\label{xfig5}
\end{figure}

Note that for  the   LSND region with
$\Delta m_{LSND}^2 = (6 - 8) \ \ {\rm eV}^2$ allowed at 99 $\%$ C.L.  one gets
$m_{\beta}> \sqrt{\Delta m_{LSND}^2}\simeq 2.5~ {\rm eV}$
which is already excluded by the Mainz result at 95 $\%$ C.L.
(see fig. \ref{xfig4}).
Hence, for the schemes with  inverted hierarchy the preferable range of
mass would be
\be
0.4 {\rm \ \ eV} < m_{\beta} <1.75 {\rm \ \ eV}.
\label{ex}
\ee

{\it 1). The (3 + 1) scheme.} The upper Bugey bound and the lower LSND
bound on $1 - \rho_e^h$ are  shown in fig. \ref{xfig4}.
They are similar to the bounds on $\rho_e^h$ in schemes with normal mass
hierarchy.

Let us consider  the implications of the $2 \beta 0 \nu$-decay searches.
The contribution of the light states  to $m_{ee}$   is
negligible. As a consequence,  the  bounds imposed by the
$2 \beta 0 \nu$-decay searches  are
similar to the bounds in the three-neutrino scheme,
and  the only difference is that in the latter scheme,
$m_{\beta}$ is a free parameter, whereas
in the (3+1) scheme it equals to  $\sqrt{\Delta m_{LSND}^2}$.

For the SMA solution of the solar neutrino problem we have
$m_{ee} \approx \sqrt{\Delta m_{LSND}^2} > 0.4$ eV which is already larger
than the present upper bound. So, keeping in mind the uncertainties of
the nuclear matrix element, we can say that this possibility is
disfavored.

In fig. \ref{xfig4}, we show the upper bounds on $m_\beta$
($m_\beta<{m_{ee}^{max} / \cos 2 \theta_\odot})$
which correspond to $|U_{e3}|$=0 and
$m_{ee}^{max}=0.34$ eV. The bounds are shown for different values of
sin$^2 2 \theta_\odot$ from the  large mixing solution regions. Note
that for the LMA solution the 99 \% upper bound, sin$^2 2\theta_\odot
\simeq$0.95,
gives $m_\beta <$ 1.5 eV, which is already below the present
kinematic bound. The best fit region  sin$^2 2\theta_\odot
\sim (0.6 - 0.8)$ leads to $m_\beta\simeq (0.55 - 0.75)$ eV well in
the range of the KATRIN sensitivity.
For the LOW solution the maximal mixing is possible for which
$\cos 2\theta_\odot = 0$, and the whole  region of $m_\beta$ up to
the present kinematic bound (\ref{mainz}) is accepted.\\

{\it 2). The (2 + 2) scheme.}
Here the effects  are similar to those in the (3 + 1) scheme with two
differences:

i). As it is shown in  fig. \ref{xfig5}, the LSND result is less
restrictive and the
rate in the tail can be substantially lower.

ii). The bound on $m_{\beta}$ from the neutrinoless double
beta decay coincides with the bound  in the (3 + 1)
scheme  at  $|U_{e3}| = 0$:
\be
m_{ee} < m_{\beta} <  \frac{m_{ee}}{\cos 2 \theta_{\odot}}~ .
\label{U=0}
\ee

\section{Non-hierarchical schemes}

Let us consider schemes in which the masses of  states from the light set
are also in the  range of sensitivity of  KATRIN experiment: $m_1 > 0.3$
eV.
Clearly, these states  are
quasi-degenerate,   and their effect on the $\beta$-decay spectrum can
 be characterized by the  effective mass, $m_{\beta}^l$, and the total
coupling with
the electron neutrino, $\rho_e^l$. From the unitarity condition we have
$\rho_e^l + \rho_e^h = 1$.

In the non-hierarchical  schemes, the effect of neutrino mass
on the $\beta$-decay spectrum
consists of the kink and the shift of the end point. The kink is at
\be
E_e^{kink} = E_0 - m_{\beta}^h,
\ee
where
\be
m_{\beta}^h  = \sqrt{\Delta m_{LSND}^2 + (m_{\beta}^{l})^2},
\label{massshift}
\ee
and its size is determined by  $\rho_e^h$. In contrast to the hierarchical
case, the mass  of the heavy set is not fixed  by $\sqrt{\Delta m_{LSND}^2}$.
The tail above the kink, $E_e > E_e^{kink}$,  is described
by $\rho_e^l$ and the end point is shifted to
\be
E_e^{ep} = E_0 - m_{\beta}^l .
\ee

Note that if $m_\beta ^l \simeq  0.5$ eV,
 from (\ref{massshift}) we obtain the effective mass of the heavy set
$m_{\beta}^h \simeq 2.5,~ 1.5$ and  $0.8$ eV for $\Delta m_{LSND}^2 = 6,~
1.5 ~$ and $0.4$ eV$^2$ correspondingly. For large values of
$\Delta m_{LSND}^2$ both the  structures (the kink and the bending  of the
spectrum at  the end point) can in principle be separately detected by
KATRIN. For $\Delta m_{LSND}^2$ at the lower allowed end
the difference of the two effective masses  becomes small:
$m_{\beta}^h - m_{\beta}^l < 0.3$ eV, and these two
structures may not be resolved.

With increase of  $m_{\beta}^l$ the scheme  transforms into the
scheme with four degenerate neutrinos. Already at  $m_{\beta}^l = 1$ eV,
we get the mass of the heavy set
$m_{\beta}^h = 2.7,~ 1.5,$ and  $1.2$ eV,
for $\Delta m_{LSND}^2 = 6,~ 1.5,~$ and $0.4$ eV$^2$.
For the  smallest   mass, $m_{\beta}^h$=1.2 eV,
the difference of masses, $m_{\beta}^h - m_{\beta}^l < 0.2$ eV,
is too small to be resolved.
With the increase of  $m_{\beta}^l$ the two structures in the spectrum,
the kink and the bending, merge.

The parameters of the kink and of the tail depend on specific properties
of the scheme. We will call it  the scheme with normal ordering
of states when the electron neutrino is distributed
mainly in the light set. And we
will refer to the opposite situation, when
$\nu_e$ is in the heavy set, as to the scheme with inverted order of states.

In the non-hierarchical schemes the mass spectrum is shifted to larger
values of mass.
The oscillation pattern, however,  is not changed
and the oscillation bounds are  the same
as in  the  hierarchical schemes
described above (see fig. \ref{xfig2}, \ref{xfig5}).
However, the implications of the  $2 \beta 0 \nu$-decay
searches are changed.

\subsection{Schemes with normal order of states}

The electron neutrino is mainly  in the light  set.
The beta decay spectrum should have a small kink at
$E_0 - m_{\beta}^h$ with the size restricted by  Bugey experiment:
$\rho_e^h < 0.027$,  and  a strong bending of spectrum at $E_0 -
m_{\beta}^l$.

\begin{figure}
\begin{center}
\vskip -3.2cm
\hskip -7.0cm
\parbox[c]{3.5in}
{\mbox{\qquad\epsfig{file=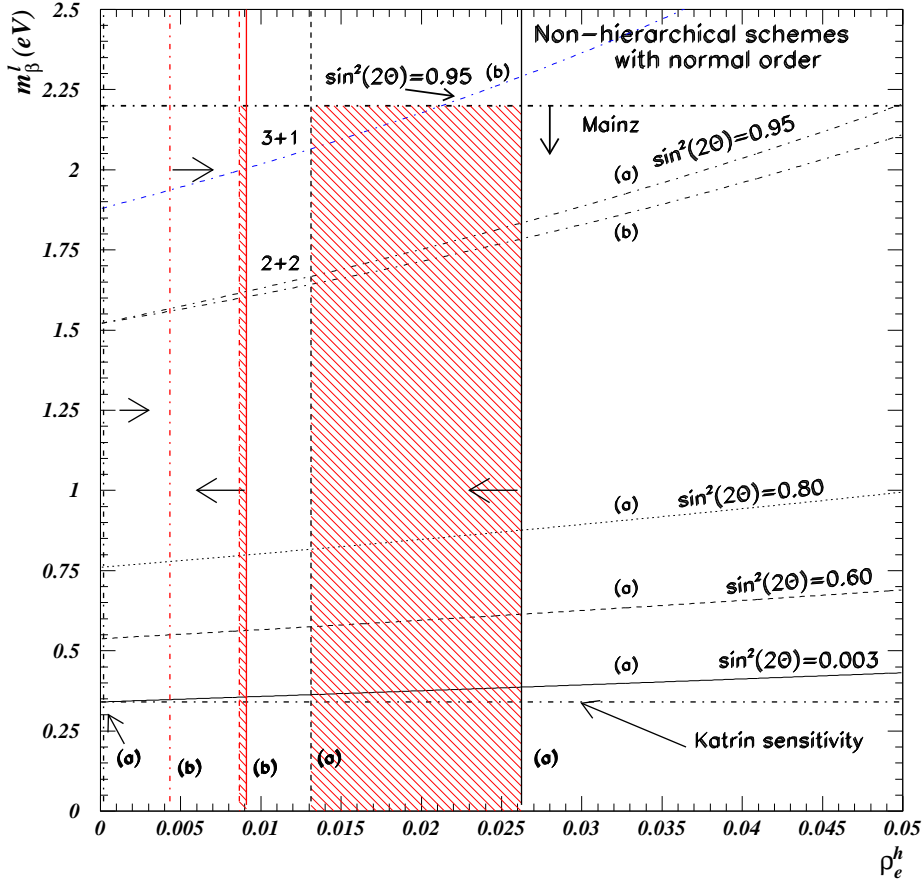,width=1.5\linewidth,height=1.5\linewidth}}}
\end{center}
\vskip -0.8cm
\caption{
The bounds on the effective  mass of the light set
$m_{\beta}^l$,  and the coupling of the electron
neutrino with heavy set,  $\rho_e^h$,
in the  non-hierarchical schemes with normal order
of levels.
The vertical solid lines  show  the upper bounds
on $\rho_e^h$
from
Bugey experiment. The dashed and dash-dotted vertical
lines show lower bounds on $ \rho_e^h$ from  LSND experiment
in the (3+1) and (2+2) schemes, respectively
(see eq. (\ref{rho_bound})). The allowed regions
for (3+1) scheme are
shadowed.
The lines with different values of  sin$^2 2 \theta_\odot$ are the
upper bounds from the
$2 \beta 0 \nu$-decay searches which correspond to $m_{ee}<$0.34 eV.
The line
denoted by ``3+1" shows the upper bound for the
(3+1) scheme with
$|U_{e3}|^2$=0.05 (see eq.
(\ref{graph2})), while the others are valid both
for the (2+2) scheme and
the (3+1) scheme with $|U_{e3}|^2$=0 (see eqs.
(\ref{graph}),
(\ref{graph2})).
The lines marked by (a) and
(b) are calculated for $\sqrt{\Delta m^2_{LSND}}$= 1.32~eV
and 0.477~eV, respectively.}
\label{xfig6}
\end{figure}

In the fig. \ref{xfig6} we show the bounds on the
$\beta$-decay
parameters
$m_{\beta}^l$ and $\rho_e^h$ for two representative values of the mass
squared difference: $\Delta m_{LSND}^2 = 1.75$ eV$^2$,
which corresponds to the weakest bound on $\rho_e^h$ from  Bugey
experiment, and $\Delta
m_{LSND}^2 = 0.227$ eV$^2$ from the lowest
(in $\Delta m^2$ scale)  LSND region.

For $\Delta m_{LSND}^2 = 1.75$ eV$^2$ the allowed region of the
$\beta$-decay parameters is between the vertical lines at $\rho_e^h = 0.013$
(dashed) and $\rho_e^h = 0.026$ (solid) for the (3 + 1) scheme
(shadowed), while for the (2 + 2) scheme  the valid
region stays between $\rho_e^h = 0.004$ (dash-dotted) and $\rho_e^h = 0.026$
(solid).

For $\Delta m_{LSND}^2 = 0.227$ eV$^2$ the allowed regions are
substantially smaller: for the (3 + 1) scheme the
region is between
lines at $\rho_e^h = 0.0095$ (dashed) and 0.010 (solid).
For the (3 + 1) scheme the region (shadowed)
is restricted by lines at 0.0002
(dash-dotted) and 0.010 (solid). All the regions are bounded from above
by the Mainz result.

The implications of the $2 \beta 0 \nu $-decay searches depend
on the specific arrangements of levels.

{\it 1. (3 + 1) scheme.}
The contribution to the effective Majorana mass, $m_{ee}$, from different
mass eigenstates can be evaluated
in the following way. The  solar pair of  states ($\nu_e$ is mainly in
these states)
yields the contribution:
\be
m_{ee}^{(1+2)} \approx m_{ee}^{sun} =  m_{\beta}^l
 (1-|U_{e3}|^2-|U_{e4}|^2)
(\cos^2 \theta_\odot  +   e^{i \delta} \sin ^2 \theta_\odot)~.
\label{bbsun}
\ee
The contributions from the two other states are
\be
m_{ee}^{(3)} = m_{\beta}^l U_{e3}^2, ~~~~ m_{ee}^{(4)} = m_{\beta}^h U_{e4}^2~.
\label{ue3}
\ee
In general these contributions may have arbitrary relative phases
and cancel each other in the  sum.
The contribution from the third level, which belongs
to the light set, is restricted by
the CHOOZ bound:  $m_{ee}^{(3)} < 0.05$ eV for $m_{\beta}^l < 1$ eV.
In turn, the fourth contribution (from the isolated level)
is restricted by the Bugey result:  $m_{ee}^{(4)} < 0.05$ eV.
The contribution  of the solar pair
(\ref{bbsun}) depends on the solution
of the solar neutrino problem and, in most of the cases, dominates
over other contributions. Indeed, for the SMA solution we get
$m_{ee}^{sun} \approx m_{\beta}^l > 0.3$  eV (if  masses of the light
set are in the
range of sensitivity of KATRIN experiment).
For the LMA solution  the cancellation of the two terms in (\ref{bbsun})
may occur
but  typically, $m_{ee}^{sun} \sim (0.2 - 1) m_{\beta}^l > 0.1$
eV is large enough to be detected in the forthcoming $2 \beta 0 \nu$-decay
experiments. For  other solutions with large  mixing
angle (LOW, VO) the cancellation can be  stronger.
The solar pair contribution can be comparable with two others
if the mixing is close to maximal:
$\sin ^2 2\theta_\odot > 0.98$ and the two terms in
(\ref{bbsun}) have opposite signs.

So, in general we expect the effective mass of the Majorana neutrino
to be $m_{ee} \sim 0.1$ eV, that is,   not too far from the
present experimental bound.
The identification of the solution of the solar neutrino problem can
clarify  a situation leading to more definite predictions.

If the solar pair gives the dominant contribution,
the measurements of $m_{ee}$,  $m_{\beta}^l$
and $\theta_\odot$ will allow to determine
the relative  phase of masses in the solar pair, $\delta$
(see eq. (\ref{bbsun})). Otherwise, due to the  presence of three
different phases we cannot determine  values of these phases, and
$m_{ee}$ will give only a lower bound on the mass scale.

Assuming  the maximal cancellation of contributions in $m_{ee}$  we
find from (\ref{bbsun}) and (\ref{ue3}),
an implicit upper bound on $m_\beta^l$ as a function of $\rho_e^h$:
\be
\left| m_\beta ^l (1- \rho_e^h) |\cos 2 \theta_\odot| - \sqrt{ (m_\beta
^l)^2+\Delta m_{LSND}^2} \rho_e^h -m_\beta ^l (1-|\cos 2 \theta_\odot | )
|U_{e3}|^2 \right| <m_{ee}.
\label{graph2}
\ee
We show this bound on $m_\beta$ for different values of $\sin^2
2\theta_\odot$ in the fig. \ref{xfig6}.
Note that  for $\sin ^2 \theta_\odot \geq$0.95, the effect of
non-zero $|U_{e3}|^2$ (the last term in the left hand side of
eq. (\ref{graph2})) is nonnegligible but it decreases with
$\sin ^2 2 \theta_\odot$. In fig. \ref{xfig6},
the line marked by ``3+1" shows the
$2 \beta 0 \nu$-decay bound for
the (3 + 1) scheme at $|U_{e3}|^2$=0.05
and $\sqrt{\Delta m_{LSND}^2}=$0.477 eV. The lower line in the pair
marked by ``2+2" is the corresponding bound for
the (3 + 1) scheme
at $|U_{e3}|^2=0$. At $|U_{e3}|^2=0$ the bounds for the
(2 + 2) and (3 + 1) schemes coincide (see below). The pair of
lines marked by
``2+2" illustrates dependence of bound on $\Delta m_{LSND}^2$. The
upper line (a) is for $\sqrt{\Delta m_{LSND}^2}=1.32$ eV while the
lower one is for  0.477 eV.
For smaller values of sin $^2 2 \theta_\odot$,
the lines have been calculated at $|U_{e3}|^2$=0 and $\sqrt{\Delta
m_{LSND}^2}$=1.32 eV. \\

{\it 2. (2 + 2) scheme.}
Now the heavy set contains two states, $\nu_3$ and $\nu_4$,
and the solar pair is in the light set.
The contribution to the Majorana mass, $m_{ee}$,  from
the light set is described by eq. (\ref{bbsun}).
The heavy  set gives
\be
m_{ee}^{(3+4)} =
m_{\beta}^h \left| |U_{e3}|^2 +
e^{i\delta'} |U_{e4}|^2 \right| < m_{\beta}^h \rho_e^h
\label{3+4}
\ee
which  is restricted by the Bugey bound:
$m_{ee}^{(3+4)} < 0.04$ eV  for  $m_{\beta}^h < 0.1$ eV.
Again, the solar pair gives the dominating contribution unless the solar
mixing
is very close to maximal
and the phase in (\ref{bbsun}) is close to $\pi$.

Assuming the  maximal cancellation of contributions in $m_{ee}$  we
find from (\ref{3+4}) and (\ref{bbsun}),
an implicit upper bound on $m_\beta^l$ as a function of $\rho_e^h$:
\be
\left| m_\beta ^l (1- \rho_e^h) |\cos 2 \theta_\odot| -
\sqrt{(m_\beta^l)^2 + \Delta m_{LSND}^2} \rho_e^h \right| < m_{ee}.
\label{graph}
\ee
We show this bound on $m_\beta$ for different values of
$\sin^2 2\theta_\odot$ in the
fig. \ref{xfig6}. The pair of lines marked by ``2+2", corresponds to
two different values of $\sqrt{ \Delta m_{LSND}^2}$:
$1.32$ eV (upper line) and $0.477$ eV (lower line).
For other values of $\sin^2 2\theta_\odot$,
$\sqrt{\Delta m_{LSND}^2}$ is taken to be $1.32$ eV. Note that
for the (3+1) scheme with $|U_{e3}|^2=0$, the bounds from
the $2 \beta 0 \nu$-decay
searches are the same as for the (2+2) scheme.

\subsection{Schemes with inverted  order of states}

The electron neutrino is mainly distributed in the
heavy set, so that one expects
a large kink at $E_0 - m_{\beta}^h$ with the size close to 1 and the
suppressed tail
with the  end point at $E_0 - m_{\beta}^l$.
In the tail the rate is restricted by the Bugey bound:
$\rho_e^l < 0.027$.
As for the case of the spectrum in the schemes with  inverted mass
hierarchy, we can
conclude that $\Delta m_{LSND}^2 < 3$ eV$^2$.

\begin{figure}
\begin{center}
\vskip -3.2cm
\hskip -7.0cm
\parbox[c]{3.5in}
{\mbox{\qquad\epsfig{file=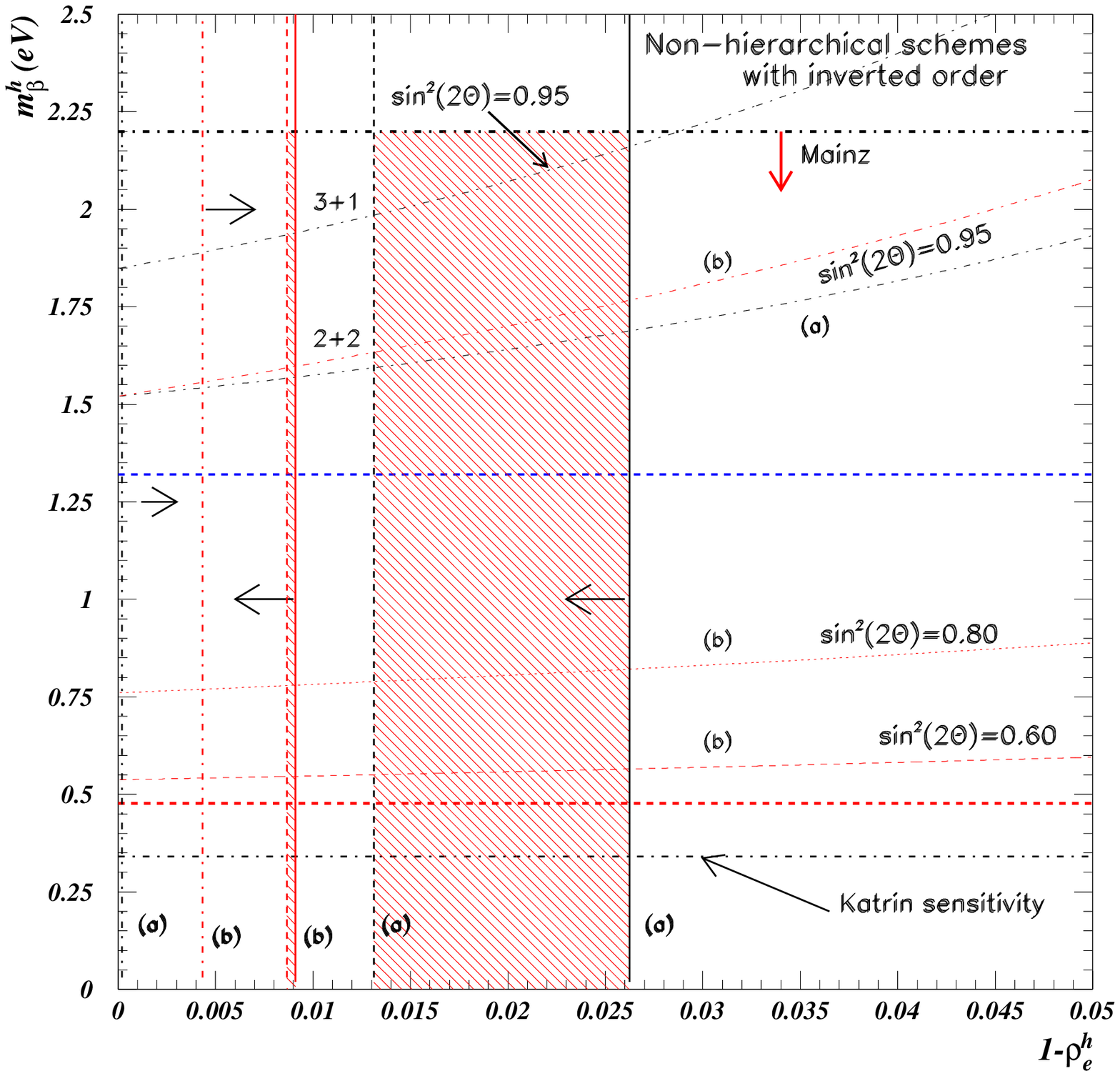,width=1.6\linewidth,height=1.5\linewidth}}}
\end{center}
\vskip -0.7cm
\caption{
The bounds on the effective  mass of the heavy set,
$m_{\beta}^h$,  and the coupling of the electron
neutrino with light set,  $1-\rho_e^h$,
in the  non-hierarchical schemes with inverted order
of states.
The vertical solid lines  show  the upper bounds
on ($1- \rho_e^h$) from
Bugey experiment. The dashed and dash-dotted vertical
lines show lower bounds on ($1- \rho_e^h$)  from  LSND experiment
in the
(3+1) and (2+2)
schemes, respectively
(see eq.
(\ref{rho_bound})). The allowed regions for the
(3+1) scheme are
shadowed.
The lines with different values of  sin$^2 2 \theta_\odot$ are the
upper bounds from the
$2 \beta 0 \nu$-decay searches which correspond to  $m_{ee}<$0.34 eV.
The line
denoted by ``3+1" corresponds to the (3+1) scheme with $|U_{e3}|^2$=0.05
(see eq.
\ref{graph3}) while others are valid both for the (2+2) scheme and
the (3+1) scheme with $|U_{e3}|^2$=0 (see eqs. (\ref{graph}),
(\ref{bbbound})).
The lines marked by (a) and
(b) are
calculated for $\sqrt{\Delta m^2_{LSND}}$= 1.32~eV
and 0.477~eV, respectively.}
\label{xfig7}
\end{figure}

In fig. \ref{xfig7}, we show the bounds on the relevant beta decay
parameters:
$m_{\beta}^h$, the effective mass of the heavy set,
and $(1 - \rho_e^h)$ which determines small admixture of
the electron neutrino in the light set. Clearly,
\be m_{\beta}^l =
\sqrt{(m_{\beta}^h)^2 - \Delta m^2_{LSND}}
\label{lightm}
\ee and
\be
m_{\beta}^h \geq \sqrt{\Delta m^2_{LSND}}.
\label{heavym}
\ee
We show  the bounds for two representative values of
$\sqrt{\Delta m_{LSND}^2}$: 1.32 and 0.447 eV
(as for the scheme with normal ordering). The allowed
ranges for $(1 - \rho_e^h)$ determined by the oscillation
experiments are the same
as the ranges of $\rho_e^h$ for normal ordering
(fig.
\ref{xfig7}). The allowed regions for the (3 + 1) scheme are
shadowed. The allowed values  of
$m_{\beta}^h$ are restricted by the Mainz limit from above and
by the LSND result from below (see eq.
(\ref{heavym})).

Let us consider the bounds from the $2 \beta 0 \nu$-decay searches.
The effective Majorana mass  can be immediately obtained
from the results for the scheme with
normal ordering by  the interchange:   $m_{\beta}^h  \leftrightarrow
m_{\beta}^l$.
This  leads to a stronger dominance of the solar pair contribution to the
effective Majorana mass.\\

{\it 1. (2 + 2) scheme.}
The contribution from the heavy set is given by
\be
m_{ee}^{h} \approx m_{ee}^{sun} = m_{\beta}^h
(1-|U_{e3}|^2-|U_{e4}|^2)
(\cos^2 \theta_\odot  +   e^{i \delta} \sin ^2 \theta_\odot),
\label{bbsolar2}
\ee
whereas the contribution from the light states  can be written as
\be
m_{ee}^{l} = m_{\beta}^l | U_{e1}^2 +  e^{i\delta'} U_{e2}^2| < m_{\beta}^l
\rho_e^l ~,
\label{bb3+4}
\ee
and according to the Bugey result: $\rho_e^l < 0.027$.
Implications of the $2\beta0\nu$ searches are similar
to those in the hierarchical case.
However now  $m_{\beta}^h$, and consequently $m_{ee}$,
can be even larger than  $\sqrt{\Delta m_{LSND}^2}$.

{} From eqs. (\ref{bbsolar2}, \ref{bb3+4}) we find the lower bound on
$m_{ee}$:
\be
m_{ee} > m_{\beta}^h  (1-\rho_e^l)|\cos 2\theta_{\odot}| -
m_{\beta}^l \rho_e^l.
\label{bbmass}
\ee
Using this inequality and the upper experimental bound on
$m_{ee}$,
we get an implicit upper bound on $m_{\beta}^h$ as a function
of $\rho_e^h$:
\be
m_{\beta}^h  \rho_e^h |\cos 2\theta_{\odot}| -
\sqrt{(m_{\beta}^h)^2 - \Delta m_{LSND}^2 }(1 - \rho_e^h)
\leq
m_{ee}^{max} = 0.34 \ \ {\rm eV}.
\label{bbbound}
\ee
The bounds for different values of the solar mixing parameter are
shown in fig.~\ref{xfig7}. The identification of the solution
of the solar neutrino problem and measurements of $\sin^2
2\theta_{\odot}$ as well as
mild improvement of the bound on the Majorana mass will have
strong impact on this scheme. For instance,
as follows from the fig.~\ref{xfig7}, the possible bounds:
$\sin^2 2\theta_{\odot} < 0.9$ and $m_{ee} < 0.1$ eV would
exclude whole the
region of parameters of the scheme down to the KATRIN
sensitivity limit.\\

{\it 2. (3 + 1) scheme.}
The contributions to $m_{ee}$ from the heavy and the light sets are equal
to:
\be
m_{ee}^{h} = m_{\beta}^h
\left[(1 - |U_{e3}|^2- |U_{e4}|^2 ) (\cos^2 \theta_\odot  +
e^{i \delta} \sin^2\theta_\odot) + U_{e3}^2
\right]
\label{bbhset}
\ee
and
\be
m_{ee}^{l} = m_{\beta}^l (1-\rho_e ^h ),
\label{bb34}
\ee
respectively. In (\ref{bbhset})
$|U_{e3}|^2$  is restricted by  the CHOOZ results
and $m_{ee}^{l} $ is restricted by the Bugey results:
$m_{ee}^{l} \stackrel
{<} {\sim} 0.03$ eV  taking $m_{\beta}^l \leq 1$ eV.
The contribution from the heavy set is similar to the one
in the scheme with three degenerate neutrinos.
For the largest part of the allowed parameter space
the contribution  of the  solar pair to $m_{ee}$  dominates.
Still significant cancellation is not excluded which
can cause  the two contributions in (\ref{bbhset})
(from the solar pair and the third mass eigenstate) to be comparable.
Note that when  $|U_{e3}|^2$ is smaller than $0.015 - 0.05$, $m_{ee}^{l}$
can be as large as $m_\beta ^h |U_{e3}|^2$.

For the SMA solution we have $m_{ee} > 0.6$ eV for $m_l > 0.3$ eV,
so that such a possbility is excluded by the present bound (\ref{m_ee}).

Assuming  the maximal cancellation of contributions in $m_{ee}$,  we
find from (\ref{bbhset}) and (\ref{bb34})
an implicit upper bound on $m_\beta^h$ as a function of $\rho_e^h$:
\be
m_\beta ^h \rho_e^h |\cos 2 \theta_\odot | -
\sqrt{(m_\beta^h)^2 - \Delta m_{LSND}^2} (1-\rho_e^h) -
m_\beta ^h (1-|\cos 2 \theta_\odot |)|U_{e3}|^2 < m_{ee}.
\label{graph3}
\ee
We show this bound on $m_\beta ^h$ for different values of $\sin^2
2\theta_\odot$ in the fig. \ref{xfig7}.
Note that for $\sin^2 2\theta_\odot$=0.95, the effect of
non-zero $|U_{e3}|^2$ is non-negligible but
for smaller values of $\sin^2 2\theta_\odot$
we can neglect $|U_{e3}|^2$.
In the fig. \ref{xfig7},
for $\sin^2 2\theta_\odot$=0.95, we
have taken $|U_{e3}|^2$=0.05 and for other values of
$\sin^2
2\theta_\odot$ we have set  $|U_{e3}|^2$ equal to zero.

\subsection{4$\nu$- schemes without LSND}

Apart from the LSND result, there is a number of other motivations to introduce
new neutrino mass eigenstates. In particular,
the sterile neutrino in the eV-range has been discussed in
connection to the supernova nucleosynthesis ($r$-processes)
\cite{r_proc}. The mixing of the keV-mass sterile neutrino
with the active neutrinos can provide a
mechanism of the pulsar kicks \cite{kicks}. The keV sterile neutrinos
may compose the warm dark matter of the Universe
\cite{hdm}. Small mixing of the sterile
neutrino with the active neutrinos can
induce the large  mixing among the active neutrinos \cite{withoutlsnd}.

Light ($SU(2)\times U(1)$)  singlet  fermions (``sterile neutrinos") can
originate from some new
sectors of the theory beyond the standard model. Neutrinos, due
to their neutrality are unique particles which can mix with these
fermions.
So, searches for  the effects of the sterile neutrinos are  of
fundamental importance even if these fermions do not solve
directly any known problem and thus their existence is not explicitly
motivated.

In this connection we will consider a general four neutrino scheme in which
three (dominantly active) neutrinos are light and the fourth neutrino
(dominantly sterile) has a mass in the eV - keV range.
The three light neutrinos may form a hierarchical structure
with the heaviest component being below 0.07 eV,
so that their masses  will
not show up in the planning beta decay experiments.
The fourth neutrino has a small mixing with active neutrinos,
and in particular, with the electron neutrino.

The scheme is similar to the (3 + 1) schemes with normal mass
hierarchy.
The difference is that now the mass $m_4$ and the mass squared differences
$m_4^2 - m_i^2$ (i = 1,2,3) are not restricted by the LSND result so that $m_4$
can be larger or much larger than 2.5 eV.

Let us consider the possible effect of this fourth neutrino in the beta decay.
We concentrate on the range of masses
$\sim$ (0.5 - 5) eV which satisfy the cosmological bounds.
The fourth state produces the kink in the beta decay spectrum
at $E = E_0 - m_4$
with the size \be \rho_e = |U_{e4}|^2. \ee Let us evaluate the
allowed range  of $\rho_e$ for different values of $m_4$. At $m_4 > 1.5$
eV the strongest bound follows from the CHOOZ result: $\rho_e < 0.027$
(we assume that other
neutrinos are much lighter, but generalization to the non-hierarchical
case is straightforward).
This
bound does not depend on the mass
for $m > 1.6$ eV.

Another direct restriction comes from the $2 \beta 0 \nu$-decay searches
provided that  neutrino is the  Majorana particle:
\be
m_4 < \frac{m_{ee}^{max}}{\rho_e},
\ee
where $m_{ee}^{max}$ is the upper bound on the effective
Majorana mass of the electron neutrino. Taking the present bound
$m_{ee}^{max} = 0.34$ eV and the maximal allowed value of $\rho_e$
we find that $m_4 < 12.6$ eV. Thus, one may see
the kink of the $3\%$ size (or smaller) in the energy interval ($0-13$)
eV below the end point.
The recent cosmological bound (\ref{recent}) shrinks substantially this
interval.

Additional restrictions on the possible effects appear if there is a substantial
admixture of the muon flavor
in the fourth state. In this case one predicts the  existence
of the $\nu_{\mu} - \nu_e$ oscillations with the effective
mixing parameter
\be \sin^2 2\theta_{e\mu} = 4 |U_{e4}|^2 |U_{\mu4}|^2
\label{newbound}
\ee
and  $\Delta m^2 \approx m_4^2 = (1 - 100)\ \  {\rm eV}^2$.
For $\Delta m^2 > 7 $ eV$^2$ the stronger bound,
$\sin^2 2\theta_{e\mu} < 1.3 \cdot 10^{-3}$,  is given by
KARMEN experiment \cite{KARMEN}.
Clearly, these bounds are satisfied,  if $|U_{\mu4}|^2$ is small
enough.
However, if $|U_{\mu4}|^2 \geq |U_{e4}|^2$ (which might be rather natural
assumption) the bound from KARMEN experiment becomes important.
Taking $|U_{\mu4}|^2 = |U_{e4}|^2$ we get from (\ref{newbound})
\be
\rho_e <  \frac{1}{2} \sqrt{\sin^2 2\theta_{e\mu}}, \label{gghh}
\ee
and for $m_4^2 > 7$ eV$^2$, it follows from (\ref{gghh}) that $\rho_e <1.7
\cdot  10^{-2}$
which is
stronger than the Bugey bound.
For heavy  neutrinos (in keV range) the neutrinoless
double beta
decay gives
very strong bound on the size of the kink: $\rho_e < 3 \cdot
10^{-4} (1
{\rm keV}/{\rm m_4})$ which will be very difficult to observe. This
bound does not exist
if the keV neutrino is the  Dirac (or pseudo-Dirac) particle.

\renewcommand{\baselinestretch}{1.2}
{\bf \section{Beta decay measurements and the neutrino mass spectrum}}
\renewcommand{\baselinestretch}{2.0}

In this section we  consider the astrophysical and cosmological
bounds on the neutrino mass. We will discuss  possible future
developments in the field.

{\subsection{Beta decay measurements and  supernova neutrinos}}

Studies of the supernova neutrinos open unique possibility to test
the schemes of neutrino mass and mixing \cite{SNspectrum}. Therefore
they may have an important impact on predictions for future beta decay
measurements.

Considering the level-crossing patterns~\cite{SNspectrum}
and the adiabaticity conditions in various resonances it is easy to show
that in all the
(2 + 2) schemes with inverted  mass hierarchy (or ordering of the states)
the originally produced $\bar{\nu}_e$-flux is
almost  completely converted
to some combination of $\bar{\nu}_{\mu}$, $\bar{\nu}_{\tau}$ and
$\bar{\nu}_s$-fluxes
at high densities in the resonance associated to $\Delta m^2_{LSND}$.
The mixing parameter in this resonance, given by
$\sin^2 2\theta_{LSND}$, is large enough to garantee
the adiabaticity of the  conversion. In this case  the  $\bar{\nu}_e$
survival
probability equals to  $P \approx \sin^2 \theta_{LSND} < 10^{-2}$.
At the same time, the $\bar{\nu}_e$-flux  observed  at the Earth
appears as a result of conversion
\be
\bar{\nu}_{\mu}, \bar{\nu}_{\tau}  \rightarrow \bar{\nu}_{e}
\label{trans}
\ee
at high densities inside the star.
Therefore the spectrum at the Earth  will practically coincide with
the hard original spectrum of $\bar{\nu}_{\mu}$ and  $\bar{\nu}_{\tau}$:
\be
F_{\bar e}(E) \approx F_{\bar{\mu}}^0(E),
\label{atearth}
\ee
and moreover, this result does not depend on the
solution of the solar
neutrino problem.  Such a  hard spectrum
of $\bar{\nu}_e$ is strongly disfavored  by the SN1987A data \cite{sn87a}.
Future detections of the Galactic supernovae can exclude the conversion
(\ref{trans}), and consequently the schemes will be excluded, completely.

In the (3+1) scheme  with inverted mass hierarchy (ordering)
the result of conversion depends on the solution of the solar neutrino
problem \cite{chance}. As in the (2+2) scheme, the original
$\bar{\nu}_e$-flux is converted to a combination of $\bar{\nu}_{\tau}$
and $\bar{\nu}_s$-fluxes.
For the SMA solution no opposite conversion (that is,
$\bar{\nu}_{\tau}$
and
$\bar{\nu}_{\mu}$ to $\bar{\nu}_e$ ) occurs. Therefore, in this scheme
the transitions lead to practically complete disappearance of the
$\bar{\nu}_e$-flux. The suppression factor, $\sin^2 \theta_{LSND} <
10^{-2}$, can not be compensated by the allowed increase
of the original  flux. The disappearance of $\bar{\nu}_e$ contradicts
the data from SN1987A, so that the scheme is excluded. Notice that this
scheme is also practically excluded by the present bound from the
neutrinoless double beta decay.

In the case of the LMA solution some part of the original
$\bar{\nu}_{\mu}-$ and $\bar{\nu}_{\tau}-$ fluxes will be transformed to
the $\bar{\nu}_e-$ flux  at low densities, so that at the surface of the
Earth one expects:
\be
F_{\bar e}(E) \approx \sin^2 \theta_{\odot} F_{\bar{\mu}}^0(E).
\label{atearth2}
\ee
Thus, $\bar{\nu}_e$ will have hard spectrum suppressed by
factor $1/3 - 1/2$. This is again disfavored by the SN1987A data.

Notice that in all these schemes the electron neutrino flux is not
elliminated  from the region proposed
for the $r$-processes~\cite{r_proc}.
So that the mechanism of production of the heavy elements
will not work.

In contrast, the  non-hierarchical  $4\nu$-schemes with
{\it normal} ordering are well consistent with the SN1987A data and
they  predict an  observable  effect in the $\beta$-decay spectrum,
as was discussed in sect. 6.

\subsection{Beta decay and forthcoming experiments}

The results of the forthcoming oscillation as well as  non-oscillation
experiments can substantially influence the both predictions of the
effects of neutrino mass and mixing in the beta decay spectrum  and
the significance of future beta decay measurements. In particular,

(i) the identification
of the solution of the $\nu_\odot$-problem and measurements of
relevant oscillation parameters,

(ii) the MiniBooNE result,

(iii) further searches for the neutrinoless double  beta decay

\noindent
will have crucial impact.
Also further improvements of the bound
on $|U_{e3}|$ will be important.  Cosmology can give a hint for
the absolute scale of neutrino mass.

Let us analyze consequences of possible results from these
experiments.

1). The  solution of the solar neutrino problem can be identified
in the forthcoming experiments:
SNO \cite{sno,mar}, KamLAND  \cite{kamland}, BOREXINO \cite{borex}.
The identification will not influence the predictions for the
$\beta$-decay immediately. Indeed, from these experiments we will
get specific values (ranges) of $\Delta m^2_{\odot}$ and
mixing angle $\theta_\odot$, {\it i.e.,}
the distribution of the electron flavor in the
solar pair of states will be determined.
But, the effects in the $\beta$-decay are not sensitive to
a particular  value of $\Delta m^2_{\odot}$, since for all possible
solutions $\Delta m_\odot ^2$ can not be resolved in $\beta$-decay
searches. Also the effects  are not sensitive to the distribution of the
electron flavor
since they are determined by the sum over states in the solar pair:
$|U_{e1}|^2 + |U_{e2}|^2 \sim 1$.
However, the identification of the solution of the solar neutrino
problem will
influence  substantially  the bounds on the $\beta$-decay effects from
the
$2 \beta 0 \nu$-decay  searches. A number of schemes discussed here
will  be excluded and for other schemes  the possible effects in the
$\beta$-decay spectrum will  be strongly restricted.

The key issues are whether the correct  solution of
the solar neutrino problem
is the small mixing
solution or the
large mixing solution,   and if it is the large mixing how large  the
deviation from maximal mixing is.

Suppose that the SMA solution will be identified, then
the following information can be obtained from
the $2 \beta 0 \nu$-decay searches in the assumption that
the Majorana neutrino exchange is the only mechanism of the decay:

\begin{itemize}

\item
For the 3$\nu$-scheme this will imply that
$m_\beta < 0.34$ eV  and further moderate improvement of the
$2 \beta 0 \nu$-decay bound will exclude the scheme.

\item
According to present data: $\sqrt{\Delta m_{LSND}^2}>0.38$ eV
(at 99 \% C.L.) \cite{LSND}. Therefore in the schemes with
inverted mass hierarchy or inverted order of states
we get $m_{ee} \geq \sqrt{\Delta m_{LSND}^2}>0.38$
eV. On the other hand,  the bound from the $2 \beta 0 \nu$-decay
is $m_{ee} < 0.34$ eV (at 90 \% C.L.). Therefore these  schemes
are excluded at  stronger than  90 \% C.L.
One should,  however, keep  in mind the uncertainties of the nuclear matrix
elements. Future double beta decay measurements will be able to
confirm  and improve the bound.
Similar conclusion can be  made for non-hierarchical schemes with
normal ordering of states and $m_l> 0.34$ eV.

Thus,  the  schemes which will survive after the identification of the SMA
solution, are the schemes
with normal hierarchy  as well as the  non-hierarchical schemes with
normal order of states and $m_l<  0.34$ eV.

\item
If $m_{ee}$ turns out to be close to the present bound
({\it e.g.}, $\sim 0.2$ eV) and the LSND result is confirmed, the only
possibility will be  the  non-hierarchical scheme with
normal ordering of states and $m_l=m_{ee}$.

\end{itemize}

Suppose now that one of the  large mixing solutions of the  solar neutrino
problem will be identified. In this case a possible cancellation
between various
contributions to  $m_{ee}$ will relax the bounds from the
$2 \beta 0 \nu$-decay.

\begin{itemize}

\item

The important point is that the  present upper bound on $m_{ee}$
is already smaller than $\sqrt{\Delta m_{LSND}^2}$:
$$
m_{ee}^{max} < \sqrt{\Delta m_{LSND}^2}
$$
(although, one should keep in mind the uncertainties of the nuclear matrix
element). At the same time,
$m_\beta ^h > \sqrt{\Delta m_{LSND}^2}$. Consequently,
the schemes with inverted mass hierarchy or inverted
ordering of states require cancellation between  the contributions
from  the solar pair states, and therefore,  there should be
CP-violating phase difference in the corresponding  mass eigenvalues.
The scheme with the SMA solution of the solar neutrino problem
is practically excluded.

The upper bound on $\sin ^2 2\theta_{\odot}$ (lower bound on the
deviation
from  maximal mixing) restricts a possible cancellation of contributions
to $m_{ee}$. This, in turn, gives an upper bound on the mass
$m_\beta ^h$. Therefore,  further improvements of the bounds on $m_{ee}$
and $\sin ^2 2\theta_{\odot}$ can strongly restrict the parameter
space of the schemes  and even exclude them.

Similar conclusions  hold for the non-hierarchical schemes
with $m_\beta ^l > m_{ee}^{max}\sim 0.34$ eV.

\item
As we discussed in section 5, the Majorana mass $m_{ee}$
gives the lower bound on the  mass of the  solar pair.
So, if $m_{ee}$ turns out to be close to the present bound (\ref{dbbound}),
we can  exclude the schemes with normal hierarchy (in which
$m_{ee} < m^l < 0.07$ eV).

\item
If $m_{ee}$ is much smaller than the present bound
({\it e.g.,} $\sim 0.01$ eV)
the scheme should have  the normal mass hierarchy or
should lead to
a strong cancellation of contributions to  the $2 \beta 0 \nu$-decay.

\end{itemize}

2).  MiniBooNE experiment will give strong discrimination
among the possibilities.

If MiniBooNE  does not confirm  the LSND result, a
large class of schemes discussed here
will be excluded.
There will  be no strong motivation to consider
the $4\nu$ schemes (see, however, sec. 6.3).

Also further improvements of  bounds on the involvement
of a  sterile neutrino in the solar
and in the atmospheric neutrino conversions may  give independent
confirmation of the $3\nu$-schemes.

We will be left with the three neutrino scheme
with strong degeneracy or with schemes having
more than three mass eigenstates without
observable signal in  MiniBooNE (see sect. 6).

So, in this case the searches of the neutrino mass in the sub-eV range
will basically test the 3$\nu$  scheme with strong degeneracy.
The observation of the  shift of end point to $E_0-m_{\beta}$ in KATRIN
experiment will be the proof of the scheme with
$m_1 \simeq m_2 \simeq m_3 \simeq m_{\beta}$.

Further insight can be obtained confronting the results of the $\beta$-decay
and the $2 \beta 0 \nu$ decay measurements, as we have discussed in
sect. 3.

If MiniBooNE  confirms the LSND result we will be
forced to consider  the $4\nu$-schemes.
In this case we get for the  absolute value of  the mass  of  heavy set:
\be
m_\nu \geq   \sqrt{\Delta m_{LSND}^2},
\ee
where the equality corresponds to
the scheme with the mass hierarchy ($m_{\beta}^h \gg m_{\beta}^l$).

  MiniBooNE experiment will not only check the LSND result but also
further restrict the oscillation parameters. Moreover,
it may  allow to disentangle the (3+1) and  (2+2) schemes.
Further searches of the sterile neutrinos in the solar
and atmospheric neutrino fluxes
should  discriminate the (2+2) and (3+1) schemes \cite{chance}.\\

3). Let us consider implications of future cosmological
measurements.
The present and future cosmological bounds on the mass scale
of neutrinos are summarized in the table \ref{limits4}.
The bound on $m_\nu$ taken from \cite{Liu}  corresponds
to  the energy density of matter $\Omega_m=0.4$ and
the reduced Hubble constant $h=0.8$.

Note that by chance the best cosmological bound (\ref{recent})
coincides numerically
with the best laboratory limit (\ref{mainz}). However, in contrast
to the latter, the cosmological bound is valid
for any flavor including the sterile neutrino,
provided that this neutrino
had been  equilibrated in the Early Universe.

The bound (\ref{recent}) (obtained for one neutrino in the eV range)
applies immediately
to the (3 + 1) scheme with normal hierarchy. In
this scheme the
heaviest (isolated)
state can produce only a small  kink in the $\beta$-decay spectrum which
will be difficult to detect. So the cosmological bound being
confronted with the value
of $\sqrt{\Delta m^2_{LSND}}$ will play important role in checking
the scheme.
The bound is even stronger for the non-hierarchical (3 + 1) schemes.

The bound is also immediately applied to the schemes with additional heavy
neutrino (sect. 6.3). Even for very small admixture of the electron
neutrino in this state ($|U_{e4}|^2 \ll 0.05$)
this, predominantly sterile, neutrino will have
equilibrium concentration in the Universe.

The bound on mass of two or three heavy degenerate neutrinos will be
stronger
than (\ref{recent}). However, the decrease  of the limit  is weaker
than just $n^{-1}$, where $n$ is the number of the degenerate neutrinos.
We estimate  the bounds for two and three neutrinos performing
rescaling
of the bound (\ref{recent}) according to results for 1, 2, and 3
neutrinos in \cite{Hu}. Thus, for two degenerate neutrinos
we get from \cite{croft} and \cite{recent} $m < 1.3$ eV.
In the  (2 + 2) schemes with normal mass hierarchy or with normal
ordering
this limit excludes the upper ``island'' of parameters allowed by  LSND
(see fig. \ref{xfig3}). In the case of inverted
hierarchy it confirms
the Mainz result.

For the three degenerate neutrinos we find $m < 0.9$ eV. This bound
excludes significant parts of the  otherwise allowed regions of the
$3\nu$-scheme and  the (3 + 1) schemes with inverted hierarchy or
order of the
states.

Forthcoming cosmological data can further substantially improve the
bounds.
In the last column of the table 1 we show the bounds which
can be obtained using  data from the Sloan Digital Sky Survey (SDSS)
\cite{Hu}
for $\Omega_m h^2 < 0.17$.

\begin{table}
\begin{center}
\renewcommand{\baselinestretch}{1.0}
\caption{Cosmological bounds on the neutrino mass scale. All the
four-neutrino schemes are considered to be hierarchical.}
\renewcommand{\baselinestretch}{2.0}
\label{limits4}
\begin{tabular}{|c|cc|cc|} \hline

 Mass scheme & present $m_{\nu} $~eV, & ref.& future $m_{\nu}
$~eV, & ref. \\
\hline
3 $\nu$ or (3+1)& 1.8 & \cite{Liu} & 0.41 & \cite{Hu}\\
inverted & 2.5 & \cite{tegmark} & & \\
\hline
(2+2) & 3.0 &  \cite{croft} & 0.57 &  \cite{Hu}\\
inverted or normal & 3.8 & \cite{tegmark} & & \\

\hline
(3+1) normal& 5.5 & \cite{croft} & 0.99 &  \cite{Hu}\\
 & 7.6 & \cite{tegmark} & & \\
 & 2.2 & \cite{recent}& & \\
\hline
\end{tabular}
\end{center}
\end{table}

\subsection{Beta decay measurements and the neutrino mass spectrum}

The identification power of the $\beta$-decay studies depends
on whether future measurements will be able to
observe the  small kink and the tail after large kink or not.

 If the $\beta$-decay measurements are not able to identify the
``small"  kinks and the suppressed tail,
the only expected distortion effect is a
shift of the end point and the corresponding bending of the spectrum.

Suppose that the LSND result will   not be confirmed and the effects of
sterile
neutrinos will not be  not found neither in the  solar, nor in
the atmospheric
neutrino fluxes. This will be the evidence for the
three neutrino scenario. (The existence of additional
sterile neutrino which mixes weakly with the block of active neutrinos
does not change the conclusion.) As we have discussed, in this case
the  $\beta$-decay parameter, $m_{\beta}$, will determine the scale
of three degenerate neutrinos.

If the LSND result is confirmed and  $\Delta m_{LSND}^2$ is measured, the
important conclusions will be drawn from the comparison of the values of
$m_{\beta}$ and $\sqrt{\Delta m_{LSND}^2}$.

Let us remind that  experiments which are insensitive to the
small kinks and tails will find
$m_{\beta} = 0$ for the schemes with normal mass hierarchy,
and $m_{\beta} = \sqrt{\Delta m_{LSND}^2}$ for  the schemes with
inverted mass hierarchy. Any type of relation is possible for the
non-hierarchical scheme with normal ordering of states:
(i) $m_{\beta} < \sqrt{\Delta m_{LSND}^2}$, (ii)
$m_{\beta} >  \sqrt{\Delta m_{LSND}^2}$ in the case of relatively small
$\Delta m_{LSND}^2$ or (iii)
$m_{\beta} \approx \sqrt{\Delta m_{LSND}^2}$ (which looks as an accidental
coincidence since this will imply the equality
$m^l = \sqrt{(m^h)^2 - (m^l)^2}$).
For the non-hierarchical schemes with inverted order one has
$m_{\beta} >  \sqrt{\Delta m_{LSND}^2}$.

Therefore:

\begin{itemize}

\item
A negative result of the neutrino mass measurements will be the evidence
of the
scheme with normal hierarchy or the
non-hierarchical  scheme  with  normal order and $m^l$  below the
sensitivity limit:
$m^l < 0.3$ eV (see eq. (\ref{futurelimit})).

\item
If it is established that
$m_{\beta} < \sqrt{\Delta m_{LSND}^2}$, the non-hierarchical scheme
with normal order of states should be selected.
Moreover, the   mass scales will be completely determined:
$m^l = m_{\beta}$, and $m^h =   \sqrt{m_{\beta}^2 + \Delta m_{LSND}^2}$.

\item
The inequality   $m_{\beta} > \sqrt{\Delta m_{LSND}^2}$
will  testify for the  non-hierarchical
scheme with inverted order of states. In this case:
$m^h = m_{\beta}$ and $m^l=\sqrt{m_\beta ^2-\Delta m_{LSND}^2}$.
The inequality can correspond also to the non-hierarchical scheme with
normal order, so that $m_{\beta} = m^l$ and
$m^h = \sqrt{m_\beta ^2 + \Delta m_{LSND}^2}$.

\item
If $m_{\beta}$  coincides (within the error bars)
with  $\sqrt{\Delta  m_{LSND}^2}$,
one of the following possibilities  will be realized :

a) the  schemes with inverted mass hierarchy  and
$m_\beta \simeq m_h \simeq \sqrt{\Delta m_{LSND}^2}$,

b) the non-hierarchical scheme with inverted order of levels and
relatively small  mass of the light set, $m^l$, so that
$m^h$ is only slightly (within the error bars) larger than $\sqrt{\Delta
m_{LSND}^2}$.

c) non-hierarchical schemes with normal order of levels and
$m_{\beta} = m^l$. In this case the equality
$m^l \sim \sqrt{\Delta m_{LSND}^2}$ is accidental.

\end{itemize}
Recently, possible
implications of results from  LSND and KATRIN experiments have
been
discussed also in \cite{pascoli}.

If  future $\beta$-decay experiments   detect  the  ``small" kinks
and the suppressed tail,
we will be  able to measure  $\rho^h_e$ and $\rho^l_e$ and unambiguously
discriminate the hierarchical schemes from the non-hierarchical ones and
the inverted scenarios from the normal ones even
without using the LSND result. This will be independent test of the
4$\nu$-scheme.
Indeed:

\begin{itemize}

\item
The spectrum with the  small kink at $m_{\beta}$ and  the tail without
bending, will be the evidence of the scheme with normal mass hierarchy.

\item
The spectrum with  the large  kink at $m_{\beta}$ and the tail without
bending will testify for  the schemes with inverted mass hierarchy.

\item
The spectrum with the  small kink and the  bending at higher energies will
correspond to the non-hierarchical scheme with normal order of the states.

\item
The spectrum with the large kink and the  suppressed tail
with shifted end point (the latter will probably be impossible
to establish)
will indicate  the non-hierarchical scheme with inverted order of states.

\end{itemize}

Notice that all these results  are the same for
the (3+1) and  (2+2)   schemes. The study of the
$\beta$-decay spectrum cannot  distinguish these schemes.
The (3+1) and  (2+2)  schemes can be discriminated
by  studies of effects of the sterile neutrinos in the
solar and atmospheric neutrino oscillations,
by  MiniBooNE experiment, by searches for the oscillations in the
$\nu_e$ and $\nu_{\mu}$ disappearance channels, etc.

The detection of the small kinks will also open a possibility to search for
the mixing of
the sterile neutrinos which are not associated with the LSND result.

\subsection{Measuring the absolute mass scale}

Without direct kinematic measurements, the absolute scale of neutrino
mass can be established only in certain exceptional cases.
This will be possible if solar neutrino data are explained  by the SMA
solution and the  $2 \beta 0 \nu$-decay is observed.
In this case $m_{ee}$ will give the mass of the solar pair:
$m^{sun} = m_{ee}$. Then using the oscillation results
one can reconstruct the whole  spectrum.

In the 3$\nu$-scheme $m_{ee}$  will immediately determine  the mass of
all the  three quasi-degenerate neutrinos.

In 4$\nu$-schemes the mass reconstruction will depend on the type of the
scheme:

In the scheme with normal mass hierarchy  $m_{ee}$ is expected to be
small: $m_{ee} \leq \sqrt{\Delta m^2_{atm}} < 0.07$ eV.
So, if a small $m_{ee}$ is detected or a strong bound on $m_{ee}$ is
obtained,
the mass of the heavy set will be given by the LSND result:
$m^h \approx \sqrt{\Delta m_{LSND}^2}$. Establishing the scale of light
masses still will be rather problematic.
Moreover, if the  $2 \beta 0 \nu$-decay searches  give only the upper bound
on
$m_{ee}$,  we should assume that
neutrinos are the Majorana particles to make conclusion on the mass.

In the non-hierarchical scheme with normal ordering  of states,  $m_{ee}$
can be as large as  the present upper bound.
Thus, $m_{ee}$ will determine the mass of the light set:
$m^l = m_{ee}$, and for the heavy set we find
$m^h = \sqrt{m_{ee}^2 + \Delta m_{LSND}^2}$.

The schemes with inverted mass hierarchy
or inverted ordering are almost excluded by the fact that
already present data indicate inequality
$(m_{ee}^{max})^2 <  \Delta m_{LSND}^2$.

For the large mixing solutions of the solar neutrino problem
(LMA, LOW, VAC) the absolute
mass scale can not be restored from $m_{ee}$
due to possible cancellation which depends on unknown
CP-violating phase.  Inversely,  the data from the $\beta$-decay
measurements
can be used to determine this phase.
If $m_{ee}$ is measured, we will be able to  put both the  lower and
the upper bounds on the absolute scale of masses:
$m_{\beta} \leq  m_{ee} / \cos 2\theta_{\odot}$ and $m_{\beta} >
m_{ee}$.

Even in those special cases where the determination
of the absolute scale is possible there are two problems:

\begin{itemize}

\item
Uncertainties of the nuclear matrix elements will
lead to significant uncertainty in the determination of  the
absolute scale.

\item
We should assume that the exchange of the light Majorana neutrinos
is the only mechanism of the $2 \beta 0 \nu$-decay.

\end{itemize}

In view of this, and keeping also in mind that the SMA solution is
disfavored by the latest solar neutrino data, we can conclude that
developments of the direct methods of determination
of the neutrino mass (and KATRIN may be only the first step)
is  unavoidable if we want to reconstruct neutrino mass spectrum
completely.

\section{Conclusions}

1. We have studied effects of the neutrino mass and mixing on the
$\beta$-decay spectrum in three neutrino schemes which explain the solar
and atmospheric neutrino data
as well as in all possible  4$\nu$-schemes which explain also the
LSND result.

We find that  schemes which can produce an observable effect
in the planned sub-eV measurements should
contain the sets (one or more) of  quasi-degenerate states.
The only exception is the (3 + 1) scheme with normal mass
hierarchy. However it  leads to a small kink which will be difficult to
observe.
We show that
the effects in the $\beta$-decay spectrum are  described by  the
effective masses  of the quasi-degenerate sets,
$m_{\beta}^{(q)}$,  and  their coupling with the electron
neutrino,  $\rho_e^{(q)}$.

2. At present, a  rather  wide class of
realistic schemes exist which can lead to an observable effect in
the sub-eV studies of the $\beta$-decay spectrum.
We show however that future oscillation experiments and $2 \beta 0 \nu
$-decay
searches can significantly restrict these possibilities.

3. The three neutrino schemes which explain the solar
and atmospheric neutrino  anomalies  in
terms of neutrino oscillations can lead to an
observable effect in  future $\beta$-decay measurements
only if all three mass eigenstates are quasi-degenerate.
The $\beta$-decay measurements give  unique possibility to identify
these schemes. Even if SMA solution is
established and the neutrinoless double beta decay is  discovered,
so that $m_{ee}$  sets the scale of the neutrino masses,
the question will rise whether the neutrino exchange
is the only  mechanism of the neutrinoless double beta decay.
Only comparison of the $m_{ee}$ with results of the beta decay measurements
will give the answer. In the case of large mixing solutions
of the solar neutrino problem,
simultaneous measurements of the $m_{ee}$
and $m_{\beta}$ as well as the solar mixing angle  open the
possibility to determine the relative CP-violating phase
of the mass eigenvalues.

4. In the four neutrino schemes which explain also the LSND
result,  observable effects in
the $\beta$-decay are strongly restricted by the Bugey and CHOOZ bounds.
Four types of
spectrum distortion are expected depending on the type of  mass hierarchy
(ordering of levels) in the scheme:

a). The spectrum with the large kink and suppressed tail above the kink.
This type of distortion realizes in the scheme with inverted  mass  hierarchy.

b). The spectrum with small kink.
This type of distortion is expected
in the scheme with normal  mass  hierarchy.

c). The spectrum with small kink and ``strong" bending at the shifted end
point. Such a distortion corresponds to the non-hierarchical spectrum with
normal ordering of states.

d). Spectrum with large kink and strongly suppressed tail above the kink
 and a shift of end point. This type of distortion realizes in
the non-hierarchical scheme with inverted   order of states.

The rates in the suppressed tails and sizes of small kinks are determined by
$\rho_e^h$ or (1$-\rho_e^h$) and the latter quantities are restricted by
the Bugey or CHOOZ bounds: $\rho_e^h\leq 0.027$.

The lower bound on $\rho_e^h$ appears from the LSND result. This bound is close
to the Bugey upper bound in the (3+1) schemes and it can be  much weaker for
the (2+2)
schemes.

The $4\nu$-schemes with inverted mass hierarchy are disfavored
by the data from SN1987A, leading  to a hard spectrum of
$\bar{\nu}_e$.

5. The identification power of the $\beta$-decay measurements
will depend  on the possibility to detect
the suppressed tail and the small-size kinks in the spectrum.

Even if small kinks or suppressed tails are unobservable,
the important conclusions can be drawn from comparison of the values of
$m_{\beta}$ and $\sqrt{\Delta m^2_{LSND}}$. This will allow
one to identify
the type of mass hierarchy (in $4\nu$-schemes) and also to distinguish
the hierarchical from non-hierarchical schemes.
Note that if  the small kink or suppressed tail are unobservable, the
effect expected from the
presently favored schemes of neutrino masses is the same as the effect of
the electron neutrino with  definite mass.

 Observations of the small kinks or suppressed tails  will allow
us to measure the mixing parameters
$\rho_e^h$ and $\rho_e^l$ and to  make the independent
identification of the scheme.
The $\beta$-decay measurements can  also distinguish the
three and four  neutrino schemes. However, this
can be done  by MiniBooNE and other experiments even before
new $\beta$-decay results will be available.

6. Even if the LSND result is not confirmed, there are some
motivations to
search for the kinks in the energy interval  ($1 - 10$) eV below the end
point.
The kinks can be due to mixing of the active neutrinos with the light
singlet
fermions which originate from some other sectors of  theory
beyond the Standard model.

7. The important conclusions can be drawn from the combined analysis  of
results of the $\beta$-decay measurements, $2\beta0\nu$-searches
and the identification of the solution of the solar neutrino
problem.

8. Negative results of future $\beta$-decay  experiments
will have a number of important implications. In particular,

- Large part of the parameter space of the 3$\nu$-schemes
with degenerate mass spectrum will be excluded;

- If the LSND result is confirmed, and  the bound from the beta decay
is  $m_{\beta} < \sqrt{\Delta m^2_{LSND}}$ the schemes with inverted
hierarchy (order) as well as the  schemes  with normal order and
$m^l > \sqrt{\Delta m^2_{LSND}}$ will be excluded.\\

If we  want eventually to know the whole ``story about neutrinos"
we should measure the absolute values of their masses.
{}From the point of view of  implications for  the fundamental theory,
for
astrophysics and cosmology, the masses are  at least as important as
other neutrino parameters such as the mixing angles and
CP-violating phases.
As follows from our study, to reconstruct the absolute values of masses
unambiguously and without additional assumptions
one needs  almost unavoidably
to develop and to perform new direct (kinematic) measurements
(or look for some new alternatives).
The KATRIN experiment may be just the first step.
There is no reason, why we should devote less time,  effort and
resources
for determination of the absolute scale of neutrino  mass
than, {\it e.g.},
we are going to devote to measurements of  the oscillation
parameters
and the CP-violating phases.

\renewcommand{\baselinestretch}{1.2}
{\bf \section*{Appendix: Effective mass of the set of quasi-degenerate states}}
\renewcommand{\baselinestretch}{2.0}

The effect of the  set of quasi-degenerate states on the $\beta$-decay spectrum
can be described by the  effective mass, $m_\beta$, and the coupling
$\rho_e$ of the set with the electron neutrino. Let us evaluate  the
accuracy of such an approximation.
The error is maximal in the energy interval  close to
$E_0-m_i$, where
$m_i$ is the mass of the lightest state from the quasi-degenerate set.
Let us compare number of events (decays) due to states from this set in
the interval:
$(E_0-m_i- \Delta E) - (E_0-m_i)$, using (1) the effective parameters
$\rho_e$ and  $m_\beta$: $\Delta n(\rho_e , m_\beta )$,  and
(2) the precise parameters of states:
$\Delta n(U_{ei}, m_i)$. Here $\Delta E$ is the energy interval which can be
resolved by the detector. Let us calculate
$$
R \equiv
{\Delta n(\rho_e,m_\beta )-\Delta n(U_{ej},m_j) \over \Delta
n(\rho_e,m_\beta )}.
$$
Since in the energy range that we are interested, $F(E,Z)$  and
$p$ in (\ref{re}) are smooth
functions of energy  we can factor out them and estimate the
relative error as follows
$$
R \simeq 1-{\sum_j |U_{ej}|^2
\int_{E_0-\Delta E -m_i}^{E_0-m_i} (E_0-E)\sqrt{(E_0-E)^2-m_j^2}
\Theta (E_0-E-m_j )dE
\over
\int_{E_0-\Delta E -m_i}^{E_0-E} \rho_e (E_0-E)
\sqrt{(E_0-E)^2-m_\beta ^2} \Theta (E_0-E-m_\beta) dE}
$$

$$={\rho_e [(m_i+\Delta E)^2-m_\beta ^2]^{3
\over 2}-\sum_j |U_{ej}|^2[(m_i+\Delta E)^2-m_j^2]^{3 \over 2} \over
\rho_e
[(m_i+\Delta E)^2-m_\beta ^2]^{3 \over 2}},
$$
where  in the sum $j$  runs over the set.

Note that
although the derivative of spectrum at $E_0-m_i$ is divergent,
$\Delta n$ has finite derivative.
If $ |m_j -m_\beta | \ll m_\beta $ one can expand the relative error around
$m_\beta$ over
$\Delta
m / ((m_i+\Delta E)^2-m_\beta ^2)$ as follows
\be
R= {3\sum_j |U_{ej}|^2 m_\beta \Delta m_{j} \over \rho_e [(m_i+\Delta
E)^2-m_\beta ^2]}+{3\sum_j |U_{ej}|^2[2 m_\beta ^2-(m_i+\Delta E)^2](\Delta
m_{j})^2 \over \rho_e [(m_i+\Delta E)^2-m_\beta ^2]^2},
\label{dm}
\ee
where $\Delta m_j \equiv m_j - m_\beta $.
Using the inequality ${\Delta m \ll \Delta E}$, we can rewrite
(\ref{dm}) as
\be \label{appbound}
R= {3\sum_j |U_{ej}|^2 m_\beta \Delta m_{j} \over \rho_e (2 m_i \Delta
E+\Delta
E^2)}+{3\sum_j |U_{ej}|^2[ m_i ^2-2 m_i\Delta E-\Delta E^2](\Delta
m_{j})^2 \over \rho_e [2 m_i \Delta E+\Delta E^2]^2}.
\ee
If  bending of the energy spectrum is observable, $\Delta E$ is of
the order of $m_i$ or smaller. Consequently, the first term in (\ref{dm}) is
of
the order  of $\Delta m / \Delta E$ and the second term is of order
$(\Delta m / \Delta E)^2$. If we set
$$
m_\beta={\sum_j m_j |U_{ej}|^2 \over \sum_j |U_{ej}|^2},
$$
the first term in (\ref{appbound}) vanishes
and $R$ will be of the  order $(\Delta m / \Delta E)^2$.

\section*{ Acknowledgements}
One of us (A. S.) would like to thank J. Beacom, S. Parke and
F. Vissani for useful discussions.
A. S. is grateful to
G. Drexlin  and C. Weinheimer, organizers of ``The International
Workshop On Neutrino Masses
in the sub-eV range" where preliminary results of this paper have been
reported \cite{smi}.
We thank   G. Drexlin for discussion of
sensitivity limits of  future beta decay experiments,
C. Lunardini, H. Pas and A. Ringwald for remarks concerning
the first version of the paper.
The work of O. P. was supported by Funda\c{c}\~ao de
do Estado de S\~ao Paulo (FAPESP), by Conselho Nacional
de Ci\^encia e
Tecnologia (CNPq) and by the European Union TMR network
ERBFMRXCT960090. O.  P. is grateful to GEFAN for
valuable discussions and useful comments.

\renewcommand{\baselinestretch}{1.0}
\newpage

\newpage

\newpage

\newpage

\newpage

\newpage
\newpage

\newpage

\end{document}